\documentclass[amsmath,amssymb,aps,nofootinbib,showkeys,superscriptaddress,showpacs,notitlepage]{revtex4-1}
\usepackage{graphicx, epsfig, amssymb} 
\usepackage{amsmath, amsfonts}
\usepackage{bm} 
\usepackage{color}
\usepackage[usenames]{xcolor}
\usepackage{hyperref}
\usepackage{siunitx}
\hypersetup{linktocpage,colorlinks=true,urlcolor=blue,linkcolor=blue,citecolor=blue}  
\usepackage{verbatim}
\usepackage[utf8]{inputenc}
\usepackage{natbib}

\newcommand{\diff}[1]{\text{d}#1}
\newcommand{\Diff}[1]{\text{D}#1}
\newcommand*{\diag}{\operatorname{diag}}

\begin{document}
 
\title{Charged Taub--NUT solution in Lovelock gravity with generalized Wheeler polynomials}

\author{Crist\'obal Corral}
\email{cristobal.corral@usach.cl}
\affiliation{Departamento de F\'isica, Universidad de Santiago de Chile,\\ Avenida Ecuador 3493, Santiago, Chile}

\author{Daniel Flores-Alfonso}
\email{daniel.flores@correo.nucleares.unam.mx}
\affiliation{Instituto de Ciencias Nucleares, Universidad Nacional Aut\'onoma de M\'exico,\\
AP 70543, Ciudad de M\'exico, 04510, Mexico}

\author{Hernando Quevedo}
\email{quevedo@nucleares.unam.mx}
\affiliation{Instituto de Ciencias Nucleares, Universidad Nacional Aut\'onoma de M\'exico,\\
AP 70543, Ciudad de M\'exico, 04510, Mexico}
\affiliation{Department of Theoretical and Nuclear Physics,\\
Kazakh National University, Almaty 050040, Kazakhstan}


\begin{abstract}
Wheeler's approach to finding exact solutions in Lovelock gravity has been predominantly applied to static spacetimes. 
This has led to a Birkhoff's theorem for arbitrary base manifolds in dimensions
higher than four. In this work, we generalize the method and apply it to a stationary metric. 
Using this perspective, we present a Taub--NUT solution in eight-dimensional Lovelock gravity coupled to Maxwell fields. 
We use the first-order formalism to integrate the equations of motion in the torsion-free sector. The Maxwell field is presented explicitly with general integration constants,
while the background metric is given implicitly in terms of a cubic algebraic equation for the metric function.
We display precisely how the NUT parameter generalizes Wheeler polynomials in a highly nontrivial manner.
\end{abstract}

\maketitle

\section{Introduction}

Higher-dimensional theories appear in different contexts of theoretical physics. For instance, an important open problem is the question about the enormous difference between the Planck and the electroweak scale. An attempt to deal with this hierarchy problem consists in considering field theories with extra spatial dimensions~\cite{ArkaniHamed:1998rs,Randall:1999ee}. For the additional dimensions, the corresponding field equations must be generalized by including higher-curvature terms in the action. These terms appear also in the renormalization approach of quantum field theory in curved spacetimes~\cite{bd82} or in the low-energy limit of string theory~\cite{gsw87}. The AdS/CFT correspondence~\cite{Maldacena1999,Gubser:1998bc,Witten:1998qj}, on the other hand, is an additional motivation to study gravity in higher dimensions, since it provides a non-perturbative approach to strongly coupled systems by means of a weakly coupled gravitational dual within an extra-dimensional spacetime. This evidence indicates that gravitational theories with extra dimensions possessing higher-order curvature terms may have important applications in the context of quantum field theory and theoretical physics, in general.

In the case of gravity, the Lanczos--Lovelock theory is the natural generalization of General Relativity (GR) in  higher dimensions~\cite{Lanczos:1938sf,Lovelock:1971yv}. The corresponding action principle is endowed with higher-curvature terms, while sharing some of the main features of GR, namely: (i)~it is invariant under local Lorentz transformations and diffeomorphisms, (ii)~it is torsion free, and (iii)~it yields second-order field equations for the metric. This theory is free of ghosts~\cite{Zumino:1985dp} and it has the same degrees of freedom as the Einstein--Hilbert action in any dimension~\cite{Henneaux:1990au}. The first non-trivial term of the Lovelock series, i.e. the Gauss--Bonnet term, appears as a low-energy correction of string theory~\cite{Zwiebach:1985uq}, modifying the field equations in dimensions higher than four. In fact, several static exact solutions have been found in this scenario~\cite{Boulware:1985wk,Banados:1993ur,Aiello:2004rz,Aiello:2005ef,Dotti:2006cp,Dotti:2007az,Cai:2006pq,Garraffo:2008hu,Dehghani:2008qr,Anabalon:2009kq,Dadhich:2012ma}, some of which have not been studied from  a thermodynamic viewpoint or any of its more recent extensions. Although in four dimensions the Gauss--Bonnet term does not contribute to the field equations since it is a topological invariant proportional to the Euler characteristic class, its inclusion becomes relevant in the regularization of conserved charges in asymptotically locally AdS spacetimes~\cite{Aros:1999id} and in the context of holographic renormalization~\cite{Miskovic:2009bm}. Moreover, the AdS/CFT correspondence has been used to impose bounds on the shear viscosity to entropy ratio for supersymmetric CFT by considering the Lanczos--Lovelock theory as its gravitational dual~\cite{Camanho:2010ru,Edelstein:2013sna}. Quantum anomalies, on the other hand, have been computed from the holographic principle in Lanczos--Lovelock gravity, showing that the Weyl and a particular non-Abelian asymptotic symmetry are broken at the quantum level on the dual CFT~\cite{Cvetkovic:2017fxa}. Remarkably, when the theory has a unique AdS vacuum, there exists a gauge fixing that leads to a finite Fefferman--Graham expansion~\cite{Banados:2006fe}. 

One aspect of higher-dimensional gravity which is interesting for the present investigation is the (non)uniqueness of static black holes~\cite{Gibbons:2002bh,Ray:2015ava}. Indeed, consider the line element
\begin{align}
 \diff{s^2} = -f(r)\diff{t^2} + \frac{\diff{r^2}}{f(r)} + r^2\diff{\Sigma^2}, \label{static}
\end{align}
where $\diff{\Sigma^2}$ is the metric of an arbitrarily chosen codimension two submanifold, henceforth
referred to as the base manifold. In fact, the only static black hole in higher-dimensional GR which is asymptotically flat is given by the Schwarzschild--Tangherlini metric, whose base manifold is a round hypersphere. However, non-asymptotically flat solutions are obtained for different base manifolds, although the field equations imply that it must be an Einstein manifold. In Eq.~\eqref{static}, the geometry of the Einstein manifold is parametrized so that its Ricci scalar coincides with that of a hypersphere with the same dimension; this fact is closely related to the higher-dimensional Birkhoff's theorem. The Lovelock version of this result also imposes conditions on the base manifold. Nevertheless, they no longer need to be Einstein manifolds and a number of new geometries come into the fold.

Returning to Eq.~\eqref{static}, when spherical symmetry is assumed, Wheeler devised an approach to determine the metric function $f$~\cite{Wheeler:1985qd}. The differential equation for the metric function is integrated in an elementary way. Remarkably, this method yields an algebraic equation for $f=f(r)$. Moreover, defining an auxiliary function by $\mathcal{F}=(1-f)/r^2$, the general result is that
\begin{equation}
 P(\mathcal{F})\equiv\sum\limits_{i=0}^p a_i \mathcal{F}(r)^i=\frac{M}{r^{D-1}}. \label{WP}
\end{equation}
This polynomial in $\mathcal{F}$ has constant coefficients $a_i$ determined by the Lovelock coupling constants, $D$ is the spacetime dimension, $p=[(D-1)/2]$ is the highest-order curvature term contributing to the field equations, and the squared brackets denote the integer part. In Eq.~\eqref{WP}, $M$ is the integration constant, which is later related to the black hole mass.
The function $P$ is what has been dubbed the Wheeler polynomial of the solution. This showcases how incredibly restrictive spherical symmetry is. A family of $p$ spacetimes are uniquely determined by the roots of Eq.~\eqref{WP}. Of course, in higher-dimensions, the exact solutions are increasingly more complex and the lack of closed form begins in eleven dimensions for Lovelock gravity. Nonetheless, some general results have been proven to hold for this set of solutions. For instance, a solution always exists for at least one value of the sign of $M$~\cite{Wheeler:1985qd}. Moreover, the extension of the asymptotic solution increases monotonically as $r$ decreases, until it ends for small values of $r$ in one of the following two possibilities: either a curvature singularity at the origin is surrounded by exactly one event horizon or the singularity happens at a finite value of $r$, where at most one event horizon is present. Notice that this includes the possibility of a naked singularity.

All maximally symmetric spaces are equally restrictive. The topological versions of these solutions
are determined by Wheeler polynomials as well, but with the auxiliary function redefined as $\mathcal{F}=(\kappa-f)/r^2$,
where $\kappa$ is the spatial sectional curvature.
However, the most general admissible base manifolds require the use of an analogue of Wheeler's polynomial
defined by~\cite{Ray:2015ava}
\begin{equation}
 Q(U)\equiv\sum\limits_{k=0}^p b_k r^{-2k}A_k(U)=\frac{M}{r^{D-1}}, \label{AWP}
\end{equation}
where the constants $b_k$ depend on the geometry of the base manifold and the auxiliary
function is defined as $U=-f/r^2$. Wheeler polynomials~\eqref{WP} are rearrangeable as 
just above, e.g., spaces of constant sectional curvature $\kappa$ have constants $b_s=\kappa^s$.
The polynomials $A_k$ are of order $p-k$ and are defined by
\begin{equation}
 A_k(U)\equiv\sum\limits_{i=0}^p \binom{i}{k} a_iU^{i-k},
\end{equation}
where the $a_i$ are, as before, the coefficients in Eq.~\eqref{WP}. Notice that the highest order polynomial
is $A_0=P(U)$ and that the polynomials comply with the recurrence relation $A'_k=(k+1)A_{k+1}$.

Even outside the context of determining exact solutions, Wheeler polynomials provide a remarkable theoretical tool
to investigate gravitational physics. Equation~\eqref{AWP}, for instance, provides a way for black hole thermodynamics
to be carried out even when a closed form for $f$ is not available~\cite{Cai:2003kt,Kastor:2011qp}. In essence, this can be carried out because black holes have event horizons characterized by the vanishing of the metric function. Hence, the Wheeler polynomial may be evaluated in the null hypersurface to yield an important algebraic relation. Taking the differentiation of the polynomial and 
restricting to the horizon determines the Hawking temperature. Of course, it is crucial to relate the integration constant with the physical parameters of the solution, especially the mass of the black hole. The relation between gravitational parameters and thermodynamical ones allows for a vast class of scenarios to be explored in this direction. However, staticity need not limit this line of research.

The original Taub and Newman--Tamburino--Unti metrics~\cite{Taub:1950ez,Newman:1963yy}---hereon Taub--NUT---have motivated a plethora of investigations in gravitational physics. A particular research area is spacetime thermodynamics, where the similarities between Taub--NUT metrics and black holes have been studied through Euclidean techniques. Relying on the methods of finite-temperature quantum field theory, an analytical continuation of the metric is performed, and the period of the Euclidean time circle is chosen in such a way that no conical singularity is present. The action of the $U(1)$ isometry group, in general, has a set of fixed points which comes from the Killing horizon in the Lorentzian sheet. If the set is zero-dimensional, the analytically continued sheet is called Taub--NUT; otherwise, it is dubbed Taub-Bolt. Possible observational signatures of this spacetimes have been studied in~\cite{Chakraborty:2017nfu,*Chakraborty:2019rna}. 

Higher-dimensional Taub--NUT and Taub-Bolt metrics are a special type of inhomogeneous geometry on complex line bundles over a K\"ahler manifold~\cite{Bais:1984xb,Page:1985bq,Taylor:1998fd}. Thus, they exist only in even dimensions. These metrics have Lorentzian counterparts which in the static limit coincide with Eq.~\eqref{static}; in this case, the base manifold is K\"ahler. In fact, Taub--NUT geometries, in Boyer--Lindquist coordinates, quite resemble the line element~\eqref{static}. This,
in turn, implies that the Wheeler approach is applicable to these stationary spacetimes as well. Of course, the method
is blind to whether the metric is Lorentzian or not.
These spacetimes carry a gravitational charge which in many ways is analogous to a magnetic monopole moment (for a recent discussion see~\cite{Flores-Alfonso:2018jra}). An important example are the famous Kaluza--Klein monopoles \cite{Sorkin:1983ns}, where the Euclidean Taub--NUT space is used as a seed manifold. Both the Taub--NUT solution and the Kaluza--Klein monopole have a rich geometric structure which have led to applications in GR~\cite{Hashemi:2018jbv,Hashemi:2018ujp} and string theory~\cite{Linshaw:2017bpf}, as well as insights in differential geometry~\cite{Li2019,Foscolo:2018mfs}. In a complementary manner, the Taub-Bolt space has a very interesting topological structure which closely resembles Euclidean black holes. This resemblance has allowed for the construction of holographic heat engines~\cite{Johnson:2017ood}. It also allows for the space to possess electromagnetic fields which generalize the Dirac monopole field~\cite{Flores-Alfonso:2017kvy}.
Taub--NUT metrics have been found to exist in a wide
range of vacuum and electrovacuum gravitational theories which include, but is not limited to, the Lanczos--Lovelock--Maxwell theory~\cite{Brill1964,Carter:1968ks,Page:1979aj,GarciaD.1984,Awad:2000gg,Awad:2005ff,Bueno:2018uoy}.

In this work, we revisit the eight-dimensional Lovelock theory where solutions in a closed form for arbitrary coupling constants are
already intractable~\cite{Hendi:2008wq}. This framework is extended by considering arbitrary coefficients for the Lovelock series and by 
adding minimally coupled Maxwell fields with general integration constants. For the sake of comparison, we use the same ans\"atze of Ref.~\cite{Dehghani:2006aa} for the metric and Maxwell fields, which can be found in Eqs.~\eqref{metricansatz}
and~\eqref{gaugeansatz}, respectively. The charged Taub--NUT solution in cubic Lovelock gravity---the main result of this work---is presented as a root of the Wheeler polynomial in Eq.~\eqref{AWP}, given by $U(r)=-f(r)/r^2$, where $f(r)$ is the metric function appearing in Eq.~\eqref{metricansatz}. The latter is determined by a generalization of Eq.~\eqref{AWP}, that is,
\begin{equation}
 Q_n(U)\equiv\sum\limits_{k=0}^p b_k r^{-2k}B_k(U)=\frac{M}{r^{D-1}}, \label{GWP}
\end{equation}
where $B_k(U)$ is a deformation of $A_k(U)$ by warping functions which depend on the NUT parameter $n$.
When the latter vanishes, we recover Eq.~\eqref{AWP}, i.e. $Q_0=Q$. Notice that in eight dimensions the polynomial is cubic, namely $p=3$. The field equations are solved by means of the first-order formalism, focusing on the torsion-free sector of the space of solutions. To the best of the authors' knowledge, this result represents the first Wheeler's-like polynomial for Taub--NUT spacetimes in Lovelock gravity.

The article is organized as follows: In Sec.~\ref{sec:lovelock} we present the eight-dimensional Lanczos--Lovelock theory coupled to Maxwell fields 
and their field equations. In Sec.~\ref{sec:wheeler}, we restrict ourselves to lower orders in the Lovelock series and write the (analogue) Wheeler polynomials
for a spherical and complex projective base manifold. This explicitly shows that, although one may freely parametrize
the base manifold to set $b_1=1$, other $b_k$ coefficients cannot be arbitrarily fixed, in contrast to the Einstein case for higher-order theories.
In Sec.~\ref{sec:taub-nut}, the higher-dimensional ansatz is presented together with lower-dimensional Taub--NUT Wheeler's polynomials which represent 
a generalization relative to Eq.~\eqref{AWP}.
In Sec.~\ref{sec:solution}, we report and discuss the charged Taub--NUT solution with arbitrary coefficients of 
the Lovelock series.Finally, conclusions and further discussions are given in Sec.~\ref{sec:conclusions}. The Appendix~\ref{sec:curvans} has been included for additional details of the computation. In our notation, greek and latin
characters denote spacetime and Lorentz indices, respectively, the Minkowski metric is $\eta_{ab} = \diag(-,+,...,+)$, and the language of differential forms will be used from hereon.

\section{Eight-dimensional Lanczos--Lovelock gravity\label{sec:lovelock}}

In this work, we use the first-order formalism to treat Lovelock's gravity~\cite{Mardones:1990qc}. This is done by considering the vielbein 
$e^a = e^{a}{}_\mu\diff{x^\mu}$ and the Lorentz connection $ \omega^{ab} = \omega^{ab}{}_\mu \diff{x^\mu}$ $1$-forms as independent gravitational fields. The former is related to the spacetime metric through $g_{\mu\nu} = \eta_{ab}e^{a}{}_\mu e^{b}{}_\nu$, where $\eta_{ab}$ is the Minkowski metric, while the latter allows us to perform the parallel transport of Lorentz-valued $p$-forms over the spacetime manifold. The curvature and torsion $2$-forms are defined through the Cartan structure equations
\begin{align}
\label{curvdef}
 R^{ab} &= \diff{\omega^{ab}} + \omega^{a}{}_c\wedge\omega^{cb} = \frac{1}{2} R^{ab}{}_{cd} e^c\wedge e^d,\\
\label{tordef}
 T^a &= \diff{e^a} + \omega^{a}{}_b\wedge e^{b} = \frac{1}{2} T^{a}{}_{bc} e^b\wedge e^c,
\end{align}
where $\wedge$ is the wedge product, $\diff{}$ is the exterior derivative, and $\Diff{}$ is the Lorentz-covariant exterior derivative with respect to $\omega^{a}{}_b$. These fields satisfy the Bianchi identities $\Diff{T^a} = R^{a}{}_b\wedge e^b$ and $\Diff{R^{ab}} = 0$. 

The eight-dimensional Lovelock theory coupled to  $U(1)$ gauge fields $A=A_\mu\diff{x^\mu}$---the theory we are interested in throughout this work---is described by the action principle
\begin{align}
 S[e^a,\omega^{ab},A] &= S_g + S_m,
\end{align}
where the gravity and matter action are denoted by $S_g$ and $S_m$, respectively, and they are considered to be minimally coupled. The Lovelock action functional is given by
\begin{align}\notag
S_g &= \int\epsilon_{abcdefgh}\bigg(\frac{\alpha_0}{8!}e^a\wedge e^b \wedge e^c\wedge e^d\wedge e^e\wedge e^f + \frac{\alpha_1}{6!} R^{ab}\wedge e^c\wedge e^d\wedge e^e\wedge e^f \\
&\quad + \frac{\alpha_2}{4!}R^{ab}\wedge R^{cd}\wedge e^e\wedge e^f + \frac{\alpha_3}{2!} R^{ab}\wedge R^{cd}\wedge R^{ef}\bigg)\wedge e^g\wedge e^h, \label{LovelockAction}
\end{align}
with $\epsilon_{01234567}=1$ (for a discussion of Lovelock gravity in terms of spacetime components see~\cite{Lovelock:1971yv}). Notice that in eight dimensions this theory admits a quartic term in the curvature $2$-form. However, it represents the dimensional continuation of the Euler density and it does not contribute to the vielbein dynamics on the bulk. The Lovelock action is conformed by a series
of dimensionally continued Euler densities. For a given dimension, the series terminates according to the differential form of maximum degree.
In addition to the gravitational sector, we write the Maxwell action functional as
\begin{align}
 S_m &= -\frac{1}{2}\int F\wedge\star F.
\end{align}
Here, $\star$ denotes the Hodge dual and $F=\diff{A}$ is the field strength of the $U(1)$ gauge fields. 

The field equations of this theory are obtained by performing stationary variations with respect to the vielbein, Lorentz connection, and $U(1)$ gauge fields, leading to 
\begin{align}\notag
0 &= \epsilon_{abcdefgh}\bigg(\frac{\alpha_0}{7!}\,e^b\wedge e^c\wedge e^d\wedge e^e\wedge e^f\wedge e^g + \frac{\alpha_1}{5!}\, R^{bc}\wedge e^d\wedge e^e\wedge e^f\wedge e^g \\
\label{eome} 
& \quad + \frac{\alpha_2}{3!}\,R^{bc}\wedge R^{de}\wedge e^f\wedge e^g + \alpha_3\, R^{bc}\wedge R^{de}\wedge R^{fg}\bigg)\wedge e^h - \tau_a, \\
\label{eomw} 0 &= \epsilon_{abcdefgh}\left(\frac{\alpha_1}{5!} e^c\wedge e^d\wedge e^e\wedge e^f +\frac{\alpha_2}{3}R^{cd}\wedge e^e\wedge e^f + 3\alpha_3 R^{cd}\wedge R^{ef} \right)\wedge T^g\wedge e^h \\
\label{eoma} 0 &= \diff{\star}F ,
\end{align}
respectively, where we have defined the energy-momentum $7$-form of the gauge fields as
\begin{align}
\label{tau}
 \tau_a &= \frac{1}{2}\Big(F\wedge\star\left(e_a\wedge F \right) - i_a F\wedge\star F\Big),
\end{align}
with $i_a$ being the inner contraction along the vector field $E_a = E^{\mu}{}_{a}{} \partial_\mu$ such that
$e^{a}{}_\mu E^{\nu}{}_{a} = \delta_\mu^\nu$ and $e^{a}{}_\mu E^{\mu}{}_{b} = \delta^a_b$. The Noether theorem associated to the invariance under diffeomorphisms~\cite{Hehl:1976kj,Hehl:1994ue,Corral:2018hxi} implies that the energy-momentum $7$-form in Eq.~\eqref{tau} satisfies the conservation law 
\begin{align}
 \Diff{\tau_a} &= i_a T^b\wedge \tau_b.
\end{align}
Invariance under local Lorentz transformations, on the other hand, imply a conservation law that is trivially satisfied for Maxwell fields. It is worth mentioning that the Bianchi identities impose severe restrictions on the torsion components 
when arbitrary coefficients of the Lovelock series are considered in vacuum~\cite{Troncoso:1999pk}. These 
restrictions can be avoided if the coefficients are chosen in such a way that the action principle can be written as the 
Chern--Simons form for the (A)dS group or as  Born--Infeld gravity in odd and even dimensions, respectively. This
implies that the theory has the maximum number of degrees of freedom~\cite{Troncoso:1999pk}.\footnote{In fact, the choice of the coefficients such that the action can be written in a Born--Infeld 
form has been used in~\cite{Hendi:2008wq} to obtain the uncharged Taub--NUT solution in third-order Lovelock theory. 
For the sake of generality, the analysis presented in this work does not assume any relation on the parameters whatsoever.}
Here, we consider arbitrary coefficients of the Lovelock series and focus our attention on the torsion-free sector of the 
space of solutions, namely $T^a=0$, which automatically solves Eq.~\eqref{eomw}. This condition allows one to solve the 
Lorentz connection in terms of the vielbein, reducing its form to the standard Levi-Civita connection. Thus, the 
solution presented here belongs to the Riemannian branch of the Lovelock theory, even though vacuum solutions with 
nontrivial torsion have been reported for different isometry groups in Refs.~\cite{Canfora:2007ux,Canfora:2007xs,Canfora:2008ka,Canfora:2010rh,Cvetkovic:2016ios,Cvetkovic:2017nkg}.

\subsection{Lower-order Wheeler polynomials\label{sec:wheeler}} 

Before going on to compute the Wheeler polynomial for the Taub--NUT solution in eight-dimensional Lovelock--Maxwell theory, it is useful to
summarize the lower-order solutions in the static limit. They portray how the original Wheeler polynomials~\cite{Wheeler:1985qd}, which consider 
spherical symmetry, are generalized to the K\"ahler case. In the next section, we explain why we specialize to the case where the base space
is complex projective. 

Let us focus on vacuum Einstein--Gauss--Bonnet theory with a cosmological constant $\Lambda$ and a Gauss--Bonnet coupling constant $\alpha_{GB}$.
This fixes the couplings constants in Eq.~\eqref{LovelockAction} in terms of these last two parameters and in particular sets $\alpha_3=0$.
In arbitrary spacetime dimension $D$, the Wheeler polynomial (\ref{WP}) is
\begin{equation}
 -\frac{2\Lambda}{(D-1)(D-2)}+\mathcal{F}+(D-3)(D-4)\alpha_{GB}\mathcal{F}^2=\frac{M}{r^{D-1}}.
\end{equation}
This equation yields the Boulware--Deser solution~\cite{Boulware:1985wk} and, setting $\alpha_{GB}=0$, leads to the familiar Schwarzschild--Tangherlini
result
\begin{equation}
 f(r)=1-\frac{M}{r^{D-3}}-\frac{2\Lambda r^2}{(D-1)(D-2)}.
\end{equation}

For comparative reasons we rewrite this result in the form of Eq.~\eqref{AWP}, which in eight dimensions is
\begin{equation}
 A_0+r^{-2}A_1+r^{-4}A_2=\frac{M}{r^7}, \label{Ray-S6}
\end{equation}
with polynomials $A_k(U)$ given by
\begin{align}
 A_0(U)&=-\frac{\Lambda}{21}+U+20\alpha_{GB} U^2, \\
 A_1(U)&=1+40\alpha_{GB} U, \\
 A_2(U)&=20\alpha_{GB},
\end{align}
recalling that $U=-f/r^2$. If we now substitute the base manifold from a hypersphere $S^6$ to a complex projective space $\mathbb{CP}^3$, then the previous polynomials remain unchanged
but the equivalent to Eq.~\eqref{Ray-S6} is
\begin{equation}
 A_0+r^{-2}A_1+2r^{-4}A_2=\frac{M}{r^7}. \label{Ray-CP3}
\end{equation}
Recall that the coefficients $b_k$ in Eq.~\eqref{AWP} depend on the geometry of the base manifold.
Since the complex projective spaces with Fubini--Study metric are Einstein manifolds, the results of Ref.~\cite{Gibbons:2002bh}
imply that Eqs.~\eqref{Ray-S6} and~\eqref{Ray-CP3} only differ in the coefficient $b_2$, once the parametrization convention of this reference is adopted. 
Additionally, we mention that
the Taub--NUT solution found in Ref.~\cite{Dehghani:2005zm} has as static limit the black hole determined by Eq.~\eqref{Ray-CP3}.
In the next section, we discuss how the NUT parameter generalizes polynomials such as the ones presented above.

\section{Higher Dimensional Taub--NUT Geometry\label{sec:taub-nut}}

The definition of a higher-dimensional Taub--NUT space we consider here is given by the family of inhomogeneous geometries built over complex line bundles
presented in~\cite{Page:1985bq}, this is
\begin{equation}\label{metricansatz}
 \diff{s^2} = f(r)\left(\diff{\tau} + 2nB\right)^2 + \frac{\diff{r^2}}{f(r)} + (r^2 - n^2)\diff{\Sigma^2},
\end{equation}
where $\tau$ is the Euclidean time coordinate and $n$ is the NUT parameter. This parameter sources the magnetic part of the Weyl tensor and it is, in general, related to the magnetic mass of the geometry~\cite{Araneda:2016iiy,Bordo:2019tyh}. Notice that, for $n\to0$, we recover a metric equivalent to~\eqref{static}, which is a static metric modulo a Wick rotation. The line element $\diff{\Sigma^2}$ is K\"ahler and its associated symplectic form is given by $\omega=\diff{B}$. The original Taub--NUT solutions are the special case where the base manifold is a sphere $S^2$, which coincides with the complex projective line $\mathbb{CP}^1$. Thus, the static limit leads to a spherically symmetric spacetime. This is particular case in four dimensions since no hypersphere admits a K\"ahler structure~\cite{Bishop:1965}. 
We will specialize to higher-dimensional Taub--NUT solutions with hyperspherical boundary conditions.
These are the only ones which admit non-singular Euclidean sheets with nuts~\cite{Bais:1984xb}.
This, in turn, implies that they are the only conditions under which Hawking--Page-like~\cite{Hawking:1982dh} phase transitions are possible~\cite{Johnson:2014pwa}.
As for Eq.~\eqref{AWP}, there is no greater loss of generality than variation of its coefficients.
These boundary conditions imply a Hopf fibration of the Euclidean time direction over a complex projective space.
Hence, we fix the geometry of the base manifold to that of Fubini--Study. For the complex projective space of real dimension $2k$
our notation is $B=\mathcal{A}_k$ and we add a subscript $k$ to the line element in (\ref{metricansatz}) to indicate that it is the
Fubini--Study metric on $\mathbb{CP}^k$.

An iterative construction of the Fubini--Study metric using explicitly real expressions is useful~\cite{Hoxha:2000jf}.
We write the recursion relation as
\begin{align}
 B &= \mathcal{A}_k = (k+1)\sin^2\psi_k\left(\diff{\phi_k} + \frac{1}{k}\mathcal{A}_{k-1}\right),\label{recpotential}\\
 \diff{\Sigma_k^2} &= 2(k+1)\bigg[\diff{\psi_k^2} + \sin^2\psi_k\cos^2\psi_k\left(\diff{\phi_k} + \frac{1}{k}\mathcal{A}_{k-1} \right)^2 
 + \frac{1}{2k}\sin^2\psi_k\diff{\Sigma_{k-1}^2} \bigg]. \label{recmetric}
 \end{align}
Notice how the metric on the $\mathbb{CP}^k$ manifold is built on top of the one on the $\mathbb{CP}^{k-1}$ submanifold.
This submanifold is in fact totally geodesic, or extrinsically flat. In these coordinates $\psi_k=\pi/2$ corresponds
exactly to this special submanifold. This fact is commented on further below. 

The four-dimensional charged Taub--NUT solution~\cite{Brill1964,Carter:1968ks} possesses a Maxwell
field whose null directions are aligned with the repeated principal null directions
of the Weyl tensor. In this spirit, we choose
\begin{equation}
 A = h(r)\left(\diff{\tau} + 2nB\right), \label{gaugeansatz}
\end{equation}
as the ansatz for the gauge potential.
This form of the gauge field was used in a higher-dimensional setting for the first time in Ref.~\cite{Awad:2005ff}. Moreover, even without an explicit form of the metric function $f$ in Eq.~\eqref{metricansatz}, we notice that the Maxwell Eq.~\eqref{eoma}
can be solved independently.
In other words, Maxwell's equations together with the ansatz~\eqref{gaugeansatz} yield a differential equation for $h$, namely
\begin{equation}
 h''\left(r^2-n^2\right)^2 + (D-2)\left[r\left(r^2-n^2\right)h' - 2n^2 h\right]=0,
\end{equation}
where prime denote derivative with respect to the coordinate $r$. This equation admits the general solution 
\begin{equation}
 h(r)=\frac{qr}{(r^2-n^2)^{k}}+\frac{vW_k}{\left(1-n^2/r^2\right)^k}, \label{generalsolutionA0}
\end{equation}
where $q$ and $v$ are integration constants and $W_k$ denotes the series
\begin{equation}
 W_k\equiv\sum\limits_{i=0}^{k}\binom{k}{i}\frac{2k-1}{2i-1}\left(-\frac{n^2}{r^2}\right)^{k-i}. \label{Wk}
\end{equation}
Notice that it resembles the binomial expansion
\begin{equation}
 \left(1-\frac{n^2}{r^2}\right)^k=\sum\limits_{i=0}^{k}\binom{k}{i}\left(-\frac{n^2}{r^2}\right)^{k-i}. \label{binom}
\end{equation}
The function $W_k$ may be generated, if so desired, by an integral formula. It may also be written in terms of Legendre polynomials or a hypergeometric function
by setting the appropriate parameters. 
To illustrate how Wheeler polynomials are generalized by NUT parameters, we present the 
special cases of Lovelock Taub--NUTs in four dimensions given by
\begin{equation}
 Q_n(U)=-\frac{\Lambda}{3}W_2+\left(1-\frac{n^2}{r^2}\right)U+r^{-2}W_1, \label{D=4}
\end{equation}
and in six dimensions by
\begin{align}
 Q_n(U)=&-\frac{\Lambda}{10}W_3+\left(1-\frac{n^2}{r^2}\right)^2U+6\alpha_{GB}W_1U^2 +r^{-2}\left[W_2+12\alpha_{GB}\left(1-\frac{n^2}{r^2}\right)U\right]
 +2r^{-4}\bigg[6\alpha_{GB}W_1\bigg]. \label{D=6}
\end{align}
Recall that $Q_n$ has been defined in Eq.~\eqref{GWP}.
Equations~\eqref{Wk} and~\eqref{binom} are, in fact, the deformation elements of the
Wheeler polynomials~\eqref{AWP} when the NUT parameter is turned on. It should be noted that, when $n\to0$, both series become unity. Recall that in four dimensions the base manifold is the complex projective line, while in six dimensions it is the complex projective plane.

\section{Charged Eight Dimensional Solution\label{sec:solution}}

We are now in a position to present the charged eight-dimensional solution which fits within the ansatz (\ref{metricansatz}). To this end, we use  
a generalized Wheeler polynomial. The base manifold is the complex projective space $\mathbb{CP}^3$.
The Euclidean time direction is Hopf fibered over this base space resulting in $r=$ constant hypersurfaces wich are hyperspheres $S^7$.
The isometry algebra of the total space is $\mathfrak{su}(4)\oplus\mathfrak{u}(1)$ and the topology will either be Euclidean, if it
has a nut, or complex projective minus a point, if it possesses a bolt.

The explicitly real Fubini--Study metric on the base manifold may be found by Eqs.~\eqref{recpotential} and~\eqref{recmetric}; thus
\begin{align}
 \mathcal{A}_1 &= 2\sin^2\psi_1\diff{\phi_1},\\
 \diff{\Sigma_1^2} &= 4 \left[\diff{\psi_1^2} + \sin^2\psi_1 \cos^2\psi_1\diff{\phi_1^2} \right],\\
 \mathcal{A}_2 &= 3\sin^2\psi_2\left(\diff{\phi_2} + \frac{\mathcal{A}_1}{2} \right),\\
 \diff{\Sigma_2^2} &= 6\left[\diff{\psi_2^2} + \sin^2\psi_2\cos^2\psi_2\left(\diff{\phi_2} + \frac{\mathcal{A}_1}{2} \right)^2 + \frac{1}{4}\sin^2\psi_2\diff{\Sigma_1^2}\right],\\
 \mathcal{A}_3 &= 4 \sin^2\psi_3\left(\diff{\phi_3} + \frac{\mathcal{A}_2}{3} \right),\\
 \diff{\Sigma_3^2} &= 8 \left[\diff{\psi_3}^2 + \sin^2\psi_3\cos^2\psi_3\left(\diff{\phi_3} + \frac{\mathcal{A}_2}{3} \right)^2 + \frac{1}{6}\sin^2\psi_3\diff{\Sigma_2^2} \right].
\end{align}
We choose the vielbein basis as shown in Appendix~\ref{sec:curvans}. 
Since we are looking for torsion-free solutions, the Lorentz connection can be solved in terms of the vielbein by solving $\diff{e^a} + \omega^{a}{}_b\wedge e^b = 0$. 
The $2$-form curvature associated with this connection can be computed from the first Cartan equation~\eqref{curvdef}. 
However, due to the cumbersome nature of its components we report them in the Appendix~\ref{sec:curvans}. 
Moreover, we write the field strength in the following manner
\begin{align}
 F &= \diff{A} = F_I e^0\wedge e^1 + F_{II}\left(e^2\wedge e^3 + e^4\wedge e^5 + e^6\wedge e^7 \right),
\end{align}
with
\begin{align}
 F_I = -h' \;\;\; \mbox{and} \;\;\; F_{II} = \frac{2n h }{r^2 - n^2}. \label{definitionFs}
\end{align}
Here, $h(r)$ is the function defined in Eq.~\eqref{gaugeansatz}. The $\mathfrak{so}(1,7)$-valued energy-momentum $7$-form~\eqref{tau} for this ansatz yields
\begin{align}
 \tau_0 &= - \rho\; e^1\wedge e^2\wedge e^3\wedge e^4\wedge e^5\wedge e^6\wedge e^7,\\
 \tau_1 &= \rho\; e^0\wedge e^2\wedge e^3\wedge e^4\wedge e^5\wedge e^6\wedge e^7,\\
 \tau_{\bar{a}} &= p\;\epsilon_{\bar{a}\bar{b}\bar{c}\bar{d}\bar{e}\bar{f}}e^0\wedge e^1\wedge e^{\bar{b}}\wedge e^{\bar{c}}\wedge e^{\bar{d}}\wedge e^{\bar{e}}\wedge e^{\bar{f}},
\end{align}
where $\bar{a}=2,..,7$ are indices of $\diff{\Sigma_3^2}$ such that $\epsilon_{234567}=1$, and
\begin{align}\label{rhopdef}
 \rho_M = \frac{F_I^2 - 3 F_{II}^2}{2} \;\;\; \mbox{and} \;\;\; p_M = \frac{F_I^2 + F_{II}^2}{2}.
\end{align}

Although we know the solution for the Maxwell field beforehand, we mention that the Maxwell equation takes the form
\begin{equation}\label{eomaans}
 F_I'\left(r^2-n^2\right) + 6\left(r F_I + nF_{II}\right)=0, 
\end{equation}
whose explicit solution is [cf. Eq.~\eqref{generalsolutionA0}]
\begin{align}
 h(r) = \frac{1}{(r^2-n^2)^3}\bigg[q r + v\left(r^6 - 5n^2 r^4 + 15n^4 r^2 + 5n^6 \right) \bigg]. \label{eightgauge}
\end{align}
In our notation, this corresponds to
\begin{align}
 F_I &= \frac{v\left(60n^6r + 40n^4r^3 - 4n^2 r^5\right) + q\left(5r^2 + n^2 \right)}{(r^2-n^2)^4}, \label{eightFI}\\
 F_{II} &= \frac{2n\left[v\left(5n^6 + 15n^4r^2 - 5n^2r^4+ r^6 \right) + qr \right]}{(r^2-n^2)^4}. \label{eightFII}
\end{align}
Examining the asymptotic behavior of the field strength reveals $q$ to be the electric charge up to some rescaling.
The other integration constant $v$ can be interpreted as the value of the electric potential at infinity~\cite{Awad:2005ff}. For the gauge potential to be regular at the nut (bolt, respectively), where the Euclidean time direction degenerates, it must be null there. So $v$ is, in fact, a potential difference
across the entire manifold. Furthermore, there is a topological interpretation of $v$ which endows it with a magnetic flavor~\cite{Flores-Alfonso:2018jra}.

This Maxwell field naturally lives in a principal $U(1)$ bundle over the Euclidean background.
The bundle's connection is locally represented by the gauge potential. This circle bundle is classified by a single topological index,
which can be calculated by integrating over the background.
If the background has a nut, then the index vanishes. In the complementary case, we have
\begin{equation}
 c=\frac{1}{16\pi^4}\int F\wedge F\wedge F\wedge F=(8nv)^4. \label{chernindex}
\end{equation}
So, we see that $v$ and $n$ are related to a topological invariant of the underlying bundle space.
However, the circle bundle just described possesses a principal $U(1)$ subbundle defined over the unique totally geodesic
sphere that lies at the asymptotic boundary. This subbundle is isomorphic to the Dirac monopole bundle and has Chern number $8nv$,
which must be an integer. In the Dirac monopole, the Chern number is twice the magnetic charge; this can be carried over to 
this eight-dimensional Taub-Bolt. For the gauge potential (\ref{eightgauge}) this means that
the magnetic charge, $p$, is given by $4nv$. This is also consistent with an asymptotic examination
such as the one carried out for the electric charge.

On the other hand, the functions $\rho_M$ and $p_M$ can be read off from Eqs. (\ref{eightFI}) and (\ref{eightFII}) by using their definition
in Eq.~\eqref{rhopdef}. Then, the field equation~\eqref{eome} reads
\begin{align}\notag
 -\rho_M &= \alpha_0 + 6\alpha_1\bigg(2R_{III} + R_{IV} + 4 R_V \bigg) + 24\alpha_2\bigg(4R_{III}R_{IV} + 16 R_{III} R_V + R_{IV}^2 \\ \notag
 &\quad + 4 R_{IV} R_V + 10 R_V^2 +  6 R_{VI}^2 \bigg) + 48\alpha_3\bigg(6 R_{III}R_{IV}^2 + 24 R_{III}R_{IV}R_V + 60 R_{III}R_V^2 \\ 
 &\quad + 36 R_{III}R_{VI}^2 + R_{IV}^3 + 6 R_{IV}R_V^2 + 18 R_{IV}R_{VI}^2 + 8 R_V^3 + 24 R_{VI}^3 \bigg),\\ \notag
 p_M &= \alpha_0 + 2\alpha_1\bigg(R_I + 10R_{III} + 2R_{IV} + 8R_V \bigg) + 8\alpha_2\bigg(2 R_I R_{IV} + 8 R_I R_V + 12 R_{II}^2 \\ \notag 
 &\quad + 20R_{III}^2 + 12 R_{III} R_{IV} + 48 R_{III} R_V + R_{IV}^2 + 4 R_{IV}R_V + 10 R_V^2 + 6 R_{VI}^2 \bigg) \\ \notag
 &\quad + 48\alpha_3\bigg(R_I R_{IV}^2 + 4 R_I R_{IV} R_V + 10 R_I R_V^2 + 6 R_I R_{VI}^2 +  12 R_{II}^2 R_{IV} + 24 R_{II}^2 R_V \\ \notag 
 &\quad + 24 R_{II}^2 R_{VI} + 12 R_{III}^2 R_{IV} + 48R_{III}^2 R_V + 2R_{III}R_{IV}^2 + 8R_{III}R_{IV}R_V + 20 R_{III}R_V^2 \\
 &\quad + 12 R_{III} R_{VI}^2 \bigg),
\end{align}
where $R_I,...,R_{VI}$ have been defined in Appendix~\ref{sec:curvans}. It is worth mentioning that these equations are not linearly independent, since differentiating the former results in the latter, after some algebraic manipulation. Thus, the equation of motion admits the following solution given in terms of a generalized 
Wheeler polynomial
\begin{equation}
 \sum\limits_{k=0}^3 b_k r^{-2k}B_k(U)=\frac{M}{r^7}+\frac{P(r)}{r^7}, \label{solution}
\end{equation}
where $M$ is an integration constant and
\begin{align}
 B_0(U)&=\frac{\alpha_0}{42}W_4+\alpha_1U\left(1-\frac{n^2}{r^2}\right)^3+20\alpha_2 U^2\left(1-\frac{2n^2}{5r^2}-\frac{3n^4}{5r^4}\right)
 +120\alpha_3U^3\left(1-\frac{n^2}{r^2}+\frac{16n^2}{5(r^2-n^2)}\right), \\
 B_1(U)&=\alpha_1W_3+40\alpha_2 U\left(1-\frac{n^2}{r^2}\right)^2+360\alpha_3U^2W_1,  \\
 B_2(U)&= 20\alpha_2W_2+360\alpha_3U\left(1-\frac{n^2}{r^2}\right), \\
 B_3(U)&= 120\alpha_3W_1. \label{Bpolys}
\end{align}
In Eq.~\eqref{solution}, the coefficients are $b_0=1,b_1=1/5,b_2=1/20$ and $b_3=1/40$. Notice that we have not set $b_1=1$ which is convenient 
in the setting of Ref.~\cite{Gibbons:2002bh}. However, it may be done so by a reparametrization of $r$. 
The left-hand side of Eq.~\eqref{solution} is completely invariant under this change except in the $b_k$ coefficients.
Moreover, $P(r)$ is the Maxwell contribution and it is a shorthand for
\begin{align}\notag
 P(r) &\equiv \frac{-1}{12r\left(r^2-n^2\right)^3}\bigg[300v^2n^{10}\left(r^2-n^2\right) + 280n^6v^2r^4\left(r^2 - 5n^2\right) - 4v^2r^8n^2\left(r^2-25n^2 \right) \\
 &\quad + 32qn^2vr^3\left(r^2-5n^2 \right) - 5q^2\left(r^2 - \frac{n^2}{5}\right)\bigg]. \label{aux}
\end{align}

To evaluate the static limit, we first interchange $v$ by its equivalent $p/4n$ and then take $n\to0$. 
After this limit has been taken, the vielbein component $e^0$ has only the Euclidean time direction. Careful evaluation yields two parts 
of the gauge potential, that we write it in the following manner
\begin{equation}
 A=\frac{q}{r^5}\diff{\tau}+2p\sin^2\psi_3\left[\diff{\phi_3} + \sin^2\psi_2\left(\diff{\phi_2} + \sin^2\psi_1\diff{\phi_1} \right) \right].
\end{equation}
The Wheeler polynomial (\ref{solution}) reduces to
\begin{equation}
 \sum\limits_{k=0}^3 b_k r^{-2k}A_k(U)=\frac{M}{r^7}+\frac{5q^2}{12r^{12}}+\frac{p^2}{48r^4},
\end{equation}
with polynomials $A_k(U)$ given by
\begin{align}
 A_0(U)&=\frac{\alpha_0}{42}+\alpha_1U+20\alpha_2 U^2+120\alpha_3U^3, \\
 A_1(U)&=\alpha_1+40\alpha_2 U+360\alpha_3U^2,  \\
 A_2(U)&= 20\alpha_2+360\alpha_3U, \\
 A_3(U)&= 120\alpha_3.
\end{align}
As far as the Wheeler polynomial is concerned, the static limit amounts to setting the warping functions in the Taub--NUT solution to unity.
Moreover, the appearance of warping functions~\eqref{Wk} and~\eqref{binom} is recurrent. The Gauss--Bonnet case ($\alpha_3=0$) shows a change of warping function from six dimensions to eight, cf. Eqs.~\eqref{D=6} and~\eqref{Bpolys}. Notice that, in eight dimensions, the coefficients that appear in the polynomials just above are recurrent in Lovelock
gravity. The cosmological constant $\alpha_0/2$ is divided by $(D-1)(D-2)/2=21$ and the Gauss--Bonnet parameter is multiplied by $(D-3)(D-4)=20$.
The factor $(D-3)(D-4)(D-5)(D-6)=120$ accompanies the cubic order coupling.

\section{Conclusions\label{sec:conclusions}}

In this work, we considered the eight-dimensional Lanczos--Lovelock-Maxwell theory in the realm of the first-order formalism of gravity. By focusing on the torsion-free sector of the space of solutions, we have generalized the Wheeler's approach of integrating the equations of motion of Lovelock theory, reducing them to an algebraic equation. This new generalization allows us to investigate stationary spacetimes, in addition to the static case which has been considered so far in the literature. In particular, we focus on Taub--NUT geometries with different higher-curvature terms of the Lovelock series to pave the way towards most general situations. The application of the method is novel since previous cases were limited only to static manifolds. Taub--NUT spacetimes are stationary and are considerably more tractable than rotating spacetimes such as the Kerr solution. Considering inhomogeneous geometries on complex line bundles over K\"ahler manifolds has proven to be a nontrivial generalization of the approaches used for static manifolds~\cite{Wheeler:1985qd,Ray:2015ava}.
However, the geometries resemble static metrics in such a way that the generalization is straightforward.

Using the extended version of Wheeler's methodology, we presented a new solution to Lanczos--Lovelock theory supplemented by Maxwell sources in a rather compact form. Arbitrary parameters of the Lovelock series are used, allowing us to analyze gravity theories such as Born--Infeld or pure Lovelock for the corresponding values of the couplings. The warping functions in the Wheeler polynomial are independent of the rescalings of the base manifold, 
except in the coefficients which encode its geometry.
The Taub-Bolt branch of the solution presented here, is a generalization of the Dirac monopole
which includes self-gravity~\cite{Flores-Alfonso:2018jra}.
It has a unique Chern index [cf. Eq.~\eqref{chernindex}] which completely classifies all possible configurations and results in an electromagnetic parameter 
being a topological charge.

Interesting questions remain open. For instance, given the recent development of Lorentzian thermodynamics for Taub--NUT spacetimes~\cite{Kubiznak:2019yiu,Ballon:2019uha,Durka:2019}, a higher-dimensional treatment including the example presented here is certainly desirable. The Euclidean method can be applied to the generalized Wheeler polynomial we provide in Eqs.~\eqref{solution} and~\eqref{aux}. We stress that this thermodynamic exploration does not require the explicit solution of the metric function, as the Wheeler polynomial suffices. The black hole limit may also deserve a thermodynamic study in the extended black hole mechanics by considering the Lovelock coupling constants as thermodynamic entities. Interpreting them as thermodynamic variables  which are held fix in the action---and so in the ensemble associated to them as well---naturally leads to their variation in the associated thermodynamic potential.
We expect to consider this task in future works.

\section*{Acknowledgments}

We thank Eloy Ay\'on-Beato, Gabriel Arenas-Henriquez, Remigiusz Durka, and Rodrigo Olea for insightful comments and helpful discussions. The work of CC is supported by Proyecto POSTDOC\_DICYT, C\'odigo 041931CM\_POSTDOC.
DFA was supported by CONACyT under Grant No. 404449. 
This work was partially supported  by UNAM-DGAPA-PAPIIT, Grant No. 111617, 
and by the Ministry of Education and Science of RK, Grant No. 
BR05236322 and AP05133630.

\appendix

\section{Vielbeins and curvature associated to an eight-dimensional Taub--NUT space\label{sec:curvans}}

For the eight-dimensional geometry we focus on, the vielbein basis has been chosen as follows
\begin{subequations}\label{ecomp}
\begin{align}
 e^0 &= \sqrt{f(r)}\left[\diff{\tau} + 8n\sin^2\psi_3\left\{\diff{\phi_3} + \sin^2\psi_2\left(\diff{\phi_2} + \sin^2\psi_1\diff{\phi_1} \right) \right\}   \right],\\
 e^1 &= \frac{\diff{r}}{\sqrt{f(r)}},\\
 e^2 &= \sqrt{8\left(r^2 - n^2\right)}\diff{\psi_3},\\
 e^3 &= \sqrt{8\left(r^2 - n^2\right)}\sin\psi_3 \cos\psi_3\left(\diff{\phi_3} + \sin^2\psi_2\left[\diff{\phi_2} + \sin^2\psi_1\diff{\phi_1} \right] \right),\\
 e^4 &= \sqrt{8\left(r^2 - n^2\right)}\sin\psi_3\diff{\psi_2},\\
 e^5 &= \sqrt{8\left(r^2 - n^2\right)}\sin\psi_3 \sin\psi_2 \cos\psi_2\left(\diff{\phi_2} + \sin^2\psi_1\diff{\phi_1} \right),\\
 e^6 &= \sqrt{8\left(r^2 - n^2\right)}\sin\psi_3 \sin\psi_2\diff{\psi_1},\\
 e^7 &= \sqrt{8\left(r^2 - n^2\right)}\sin\psi_3 \sin\psi_2 \sin\psi_1 \cos\psi_1\diff{\phi_1}.
\end{align}
 \end{subequations}
We write them here explicitly to complement recursive definitions in the main text. These recursive relations appear because
the base manifold of complex line bundle, where the metric is supported, has as base manifold a complex projective space of real six dimensions.
As is probably anticipated by the reader the geometry is that of Fubini--Study, up to a rescaling.

The components of the curvature two-form are
\begin{align}
 R^{01} &= R_I e^0\wedge e^1 + 2 R_{II} \left(e^2\wedge e^3 + e^4\wedge e^5 + e^6\wedge e^7\right),\\
 R^{02} &= R_{III} e^0\wedge e^2 +  R_{II} e^1\wedge e^3, \;\;\; R^{03} = R_{III} e^0\wedge e^3 -  R_{II} e^1\wedge e^2, \\ 
 R^{04} &= R_{III} e^0\wedge e^4 +  R_{II} e^1\wedge e^5, \;\;\; R^{05} = R_{III} e^0\wedge e^5 -  R_{II} e^1\wedge e^4,\\
 R^{06} &= R_{III} e^0\wedge e^6 +  R_{II} e^1\wedge e^7, \;\;\; R^{07} = R_{III} e^0\wedge e^7 -  R_{II} e^1\wedge e^6,\\
 R^{12} &= R_{III} e^1\wedge e^2 -  R_{II} e^0\wedge e^3, \;\;\; R^{13} = R_{III} e^1\wedge e^3 +  R_{II} e^0\wedge e^2,\\
 R^{14} &= R_{III} e^1\wedge e^4 -  R_{II} e^0\wedge e^5, \;\;\; R^{15} = R_{III} e^1\wedge e^5 +  R_{II} e^0\wedge e^4,\\
 R^{16} &= R_{III} e^1\wedge e^6 -  R_{II} e^0\wedge e^7, \;\;\; R^{17} = R_{III} e^1\wedge e^7 +  R_{II} e^0\wedge e^6,\\
 R^{23} &= 2 R_{II} e^0\wedge e^1 +  R_{IV} e^2\wedge e^3 + 2 R_{VI}\left(e^4\wedge e^5 + e^6\wedge e^7\right),\\
 R^{24} &= R_V e^2\wedge e^4 + R_{VI} e^3\wedge e^5, \;\;\; R^{25} = R_V e^2\wedge e^5 - R_{VI} e^3\wedge e^4,\\
 R^{26} &= R_V e^2\wedge e^6 + R_{VI} e^3\wedge e^7, \;\;\; R^{27} = R_V e^2\wedge e^7 - R_{VI} e^3\wedge e^6,\\
 R^{34} &= R_V e^3\wedge e^4 - R_{VI} e^2\wedge e^5, \;\;\; R^{35} = R_V e^3\wedge e^5 + R_{VI} e^2\wedge e^4,\\
 R^{36} &= R_V e^3\wedge e^6 - R_{VI} e^2\wedge e^7, \;\;\; R^{37} = R_V e^3\wedge e^7 + R_{VI} e^2\wedge e^6,\\
 R^{45} &= 2 R_{II} e^0\wedge e^1 + 2 R_{VI} e^2\wedge e^3 + R_{IV} e^4\wedge e^5 + 2 R_{VI} e^6\wedge e^7,\\
 R^{46} &= R_V e^4\wedge e^6 + R_{VI} e^5\wedge e^7, \;\;\; R^{47} = R_V e^4\wedge e^7 - R_{VI} e^5\wedge e^6,\\
 R^{56} &= R_V e^5\wedge e^6 - R_{VI} e^4\wedge e^7, \;\;\; R^{57} = R_V e^5\wedge e^7 + R_{VI} e^4\wedge e^6,\\
 R^{67} &= 2 R_{II} e^0\wedge e^1 + 2 R_{VI} e^2\wedge e^3 + 2 R_{VI} e^4\wedge e^5 + R_{IV} e^6\wedge e^7.
\end{align}
Here we have introduced various short hands, $R_I\dots R_{VI}$, which are detailed below
\begin{align}
 R_I &= -\frac{f''}{2}, &
 R_{II} &=\frac{n}{2} \frac{\diff{}}{\diff{r}}\left[\frac{f}{(r^2 - n^2)} \right], &
 R_{III} &= - \frac{f'r}{2\left(r^2-n^2\right)} +  \frac{f n^2}{\left(r^2-n^2 \right)^2}, \\
 R_{IV} &=  \frac{1}{2}\frac{1}{r^2 - n^2} - f\frac{r^2+3n^2}{(r^2-n^2)^2},&
 R_V &= \frac{1}{8}\frac{1}{r^2-n^2} - \frac{fr^2}{(r^2-n^2)^2}, &
 R_{VI} &= \frac{1}{8}\frac{1}{r^2-n^2} - \frac{fn^2}{(r^2-n^2)^2}.
\end{align}

\bibliography{references}

\begin{thebibliography}{80}%
\makeatletter
\providecommand \@ifxundefined [1]{%
 \@ifx{#1\undefined}
}%
\providecommand \@ifnum [1]{%
 \ifnum #1\expandafter \@firstoftwo
 \else \expandafter \@secondoftwo
 \fi
}%
\providecommand \@ifx [1]{%
 \ifx #1\expandafter \@firstoftwo
 \else \expandafter \@secondoftwo
 \fi
}%
\providecommand \natexlab [1]{#1}%
\providecommand \enquote  [1]{``#1''}%
\providecommand \bibnamefont  [1]{#1}%
\providecommand \bibfnamefont [1]{#1}%
\providecommand \citenamefont [1]{#1}%
\providecommand \href@noop [0]{\@secondoftwo}%
\providecommand \href [0]{\begingroup \@sanitize@url \@href}%
\providecommand \@href[1]{\@@startlink{#1}\@@href}%
\providecommand \@@href[1]{\endgroup#1\@@endlink}%
\providecommand \@sanitize@url [0]{\catcode `\\12\catcode `\$12\catcode
  `\&12\catcode `\#12\catcode `\^12\catcode `\_12\catcode `\%12\relax}%
\providecommand \@@startlink[1]{}%
\providecommand \@@endlink[0]{}%
\providecommand \url  [0]{\begingroup\@sanitize@url \@url }%
\providecommand \@url [1]{\endgroup\@href {#1}{\urlprefix }}%
\providecommand \urlprefix  [0]{URL }%
\providecommand \Eprint [0]{\href }%
\providecommand \doibase [0]{http://dx.doi.org/}%
\providecommand \selectlanguage [0]{\@gobble}%
\providecommand \bibinfo  [0]{\@secondoftwo}%
\providecommand \bibfield  [0]{\@secondoftwo}%
\providecommand \translation [1]{[#1]}%
\providecommand \BibitemOpen [0]{}%
\providecommand \bibitemStop [0]{}%
\providecommand \bibitemNoStop [0]{.\EOS\space}%
\providecommand \EOS [0]{\spacefactor3000\relax}%
\providecommand \BibitemShut  [1]{\csname bibitem#1\endcsname}%
\let\auto@bib@innerbib\@empty
\bibitem [{\citenamefont {Arkani-Hamed}\ \emph {et~al.}(1998)\citenamefont
  {Arkani-Hamed}, \citenamefont {Dimopoulos},\ and\ \citenamefont
  {Dvali}}]{ArkaniHamed:1998rs}%
  \BibitemOpen
  \bibfield  {author} {\bibinfo {author} {\bibfnamefont {N.}~\bibnamefont
  {Arkani-Hamed}}, \bibinfo {author} {\bibfnamefont {S.}~\bibnamefont
  {Dimopoulos}}, \ and\ \bibinfo {author} {\bibfnamefont {G.~R.}\ \bibnamefont
  {Dvali}},\ }\href {\doibase 10.1016/S0370-2693(98)00466-3} {\bibfield
  {journal} {\bibinfo  {journal} {Phys. Lett.}\ }\textbf {\bibinfo {volume}
  {B429}},\ \bibinfo {pages} {263} (\bibinfo {year} {1998})}\BibitemShut
  {NoStop}%
\bibitem [{\citenamefont {Randall}\ and\ \citenamefont
  {Sundrum}(1999)}]{Randall:1999ee}%
  \BibitemOpen
  \bibfield  {author} {\bibinfo {author} {\bibfnamefont {L.}~\bibnamefont
  {Randall}}\ and\ \bibinfo {author} {\bibfnamefont {R.}~\bibnamefont
  {Sundrum}},\ }\href {\doibase 10.1103/PhysRevLett.83.3370} {\bibfield
  {journal} {\bibinfo  {journal} {Phys. Rev. Lett.}\ }\textbf {\bibinfo
  {volume} {83}},\ \bibinfo {pages} {3370} (\bibinfo {year}
  {1999})}\BibitemShut {NoStop}%
\bibitem [{\citenamefont {Birrell}\ and\ \citenamefont {Davies}(1984)}]{bd82}%
  \BibitemOpen
  \bibfield  {author} {\bibinfo {author} {\bibfnamefont {N.~D.}\ \bibnamefont
  {Birrell}}\ and\ \bibinfo {author} {\bibfnamefont {P.~C.~W.}\ \bibnamefont
  {Davies}},\ }\href {\doibase 10.1017/CBO9780511622632} {\emph {\bibinfo
  {title} {{Quantum Fields in Curved Space}}}}\ (\bibinfo  {publisher}
  {Cambridge Univ. Press},\ \bibinfo {year} {1984})\BibitemShut {NoStop}%
\bibitem [{\citenamefont {Green}\ \emph {et~al.}(1988)\citenamefont {Green},
  \citenamefont {Schwarz},\ and\ \citenamefont {Witten}}]{gsw87}%
  \BibitemOpen
  \bibfield  {author} {\bibinfo {author} {\bibfnamefont {M.~B.}\ \bibnamefont
  {Green}}, \bibinfo {author} {\bibfnamefont {J.~H.}\ \bibnamefont {Schwarz}},
  \ and\ \bibinfo {author} {\bibfnamefont {E.}~\bibnamefont {Witten}},\
  }\href@noop {} {\emph {\bibinfo {title} {{Superstring theory}}}},\ Cambridge
  Monographs on Mathematical Physics\ (\bibinfo {year} {1988})\BibitemShut
  {NoStop}%
\bibitem [{\citenamefont {Maldacena}(1999)}]{Maldacena1999}%
  \BibitemOpen
  \bibfield  {author} {\bibinfo {author} {\bibfnamefont {J.}~\bibnamefont
  {Maldacena}},\ }\href {\doibase 10.1023/A:1026654312961} {\bibfield
  {journal} {\bibinfo  {journal} {International Journal of Theoretical
  Physics}\ }\textbf {\bibinfo {volume} {38}},\ \bibinfo {pages} {1113}
  (\bibinfo {year} {1999})}\BibitemShut {NoStop}%
\bibitem [{\citenamefont {Gubser}\ \emph {et~al.}(1998)\citenamefont {Gubser},
  \citenamefont {Klebanov},\ and\ \citenamefont {Polyakov}}]{Gubser:1998bc}%
  \BibitemOpen
  \bibfield  {author} {\bibinfo {author} {\bibfnamefont {S.~S.}\ \bibnamefont
  {Gubser}}, \bibinfo {author} {\bibfnamefont {I.~R.}\ \bibnamefont
  {Klebanov}}, \ and\ \bibinfo {author} {\bibfnamefont {A.~M.}\ \bibnamefont
  {Polyakov}},\ }\href {\doibase 10.1016/S0370-2693(98)00377-3} {\bibfield
  {journal} {\bibinfo  {journal} {Phys. Lett.}\ }\textbf {\bibinfo {volume}
  {B428}},\ \bibinfo {pages} {105} (\bibinfo {year} {1998})}\BibitemShut
  {NoStop}%
\bibitem [{\citenamefont {Witten}(1998)}]{Witten:1998qj}%
  \BibitemOpen
  \bibfield  {author} {\bibinfo {author} {\bibfnamefont {E.}~\bibnamefont
  {Witten}},\ }\href {\doibase 10.4310/ATMP.1998.v2.n2.a2} {\bibfield
  {journal} {\bibinfo  {journal} {Adv. Theor. Math. Phys.}\ }\textbf {\bibinfo
  {volume} {2}},\ \bibinfo {pages} {253} (\bibinfo {year} {1998})}\BibitemShut
  {NoStop}%
\bibitem [{\citenamefont {Lanczos}(1938)}]{Lanczos:1938sf}%
  \BibitemOpen
  \bibfield  {author} {\bibinfo {author} {\bibfnamefont {C.}~\bibnamefont
  {Lanczos}},\ }\href {\doibase 10.2307/1968467} {\bibfield  {journal}
  {\bibinfo  {journal} {Annals Math.}\ }\textbf {\bibinfo {volume} {39}},\
  \bibinfo {pages} {842} (\bibinfo {year} {1938})}\BibitemShut {NoStop}%
\bibitem [{\citenamefont {Lovelock}(1971)}]{Lovelock:1971yv}%
  \BibitemOpen
  \bibfield  {author} {\bibinfo {author} {\bibfnamefont {D.}~\bibnamefont
  {Lovelock}},\ }\href {\doibase 10.1063/1.1665613} {\bibfield  {journal}
  {\bibinfo  {journal} {J. Math. Phys.}\ }\textbf {\bibinfo {volume} {12}},\
  \bibinfo {pages} {498} (\bibinfo {year} {1971})}\BibitemShut {NoStop}%
\bibitem [{\citenamefont {Zumino}(1986)}]{Zumino:1985dp}%
  \BibitemOpen
  \bibfield  {author} {\bibinfo {author} {\bibfnamefont {B.}~\bibnamefont
  {Zumino}},\ }\href {\doibase 10.1016/0370-1573(86)90076-1} {\bibfield
  {journal} {\bibinfo  {journal} {Phys. Rept.}\ }\textbf {\bibinfo {volume}
  {137}},\ \bibinfo {pages} {109} (\bibinfo {year} {1986})}\BibitemShut
  {NoStop}%
\bibitem [{\citenamefont {Henneaux}\ \emph {et~al.}(1990)\citenamefont
  {Henneaux}, \citenamefont {Teitelboim},\ and\ \citenamefont
  {Zanelli}}]{Henneaux:1990au}%
  \BibitemOpen
  \bibfield  {author} {\bibinfo {author} {\bibfnamefont {M.}~\bibnamefont
  {Henneaux}}, \bibinfo {author} {\bibfnamefont {C.}~\bibnamefont
  {Teitelboim}}, \ and\ \bibinfo {author} {\bibfnamefont {J.}~\bibnamefont
  {Zanelli}},\ }\href {\doibase 10.1016/0550-3213(90)90034-B} {\bibfield
  {journal} {\bibinfo  {journal} {Nucl. Phys.}\ }\textbf {\bibinfo {volume}
  {B332}},\ \bibinfo {pages} {169} (\bibinfo {year} {1990})}\BibitemShut
  {NoStop}%
\bibitem [{\citenamefont {Zwiebach}(1985)}]{Zwiebach:1985uq}%
  \BibitemOpen
  \bibfield  {author} {\bibinfo {author} {\bibfnamefont {B.}~\bibnamefont
  {Zwiebach}},\ }\href {\doibase 10.1016/0370-2693(85)91616-8} {\bibfield
  {journal} {\bibinfo  {journal} {Phys. Lett.}\ }\textbf {\bibinfo {volume}
  {156B}},\ \bibinfo {pages} {315} (\bibinfo {year} {1985})}\BibitemShut
  {NoStop}%
\bibitem [{\citenamefont {Boulware}\ and\ \citenamefont
  {Deser}(1985)}]{Boulware:1985wk}%
  \BibitemOpen
  \bibfield  {author} {\bibinfo {author} {\bibfnamefont {D.~G.}\ \bibnamefont
  {Boulware}}\ and\ \bibinfo {author} {\bibfnamefont {S.}~\bibnamefont
  {Deser}},\ }\href {\doibase 10.1103/PhysRevLett.55.2656} {\bibfield
  {journal} {\bibinfo  {journal} {Phys. Rev. Lett.}\ }\textbf {\bibinfo
  {volume} {55}},\ \bibinfo {pages} {2656} (\bibinfo {year}
  {1985})}\BibitemShut {NoStop}%
\bibitem [{\citenamefont {Banados}\ \emph {et~al.}(1994)\citenamefont
  {Banados}, \citenamefont {Teitelboim},\ and\ \citenamefont
  {Zanelli}}]{Banados:1993ur}%
  \BibitemOpen
  \bibfield  {author} {\bibinfo {author} {\bibfnamefont {M.}~\bibnamefont
  {Banados}}, \bibinfo {author} {\bibfnamefont {C.}~\bibnamefont {Teitelboim}},
  \ and\ \bibinfo {author} {\bibfnamefont {J.}~\bibnamefont {Zanelli}},\ }\href
  {\doibase 10.1103/PhysRevD.49.975} {\bibfield  {journal} {\bibinfo  {journal}
  {Phys. Rev.}\ }\textbf {\bibinfo {volume} {D49}},\ \bibinfo {pages} {975}
  (\bibinfo {year} {1994})}\BibitemShut {NoStop}%
\bibitem [{\citenamefont {Aiello}\ \emph {et~al.}(2004)\citenamefont {Aiello},
  \citenamefont {Ferraro},\ and\ \citenamefont {Giribet}}]{Aiello:2004rz}%
  \BibitemOpen
  \bibfield  {author} {\bibinfo {author} {\bibfnamefont {M.}~\bibnamefont
  {Aiello}}, \bibinfo {author} {\bibfnamefont {R.}~\bibnamefont {Ferraro}}, \
  and\ \bibinfo {author} {\bibfnamefont {G.}~\bibnamefont {Giribet}},\ }\href
  {\doibase 10.1103/PhysRevD.70.104014} {\bibfield  {journal} {\bibinfo
  {journal} {Phys. Rev.}\ }\textbf {\bibinfo {volume} {D70}},\ \bibinfo {pages}
  {104014} (\bibinfo {year} {2004})}\BibitemShut {NoStop}%
\bibitem [{\citenamefont {Aiello}\ \emph {et~al.}(2005)\citenamefont {Aiello},
  \citenamefont {Ferraro},\ and\ \citenamefont {Giribet}}]{Aiello:2005ef}%
  \BibitemOpen
  \bibfield  {author} {\bibinfo {author} {\bibfnamefont {M.}~\bibnamefont
  {Aiello}}, \bibinfo {author} {\bibfnamefont {R.}~\bibnamefont {Ferraro}}, \
  and\ \bibinfo {author} {\bibfnamefont {G.}~\bibnamefont {Giribet}},\ }\href
  {\doibase 10.1088/0264-9381/22/13/004} {\bibfield  {journal} {\bibinfo
  {journal} {Class. Quant. Grav.}\ }\textbf {\bibinfo {volume} {22}},\ \bibinfo
  {pages} {2579} (\bibinfo {year} {2005})}\BibitemShut {NoStop}%
\bibitem [{\citenamefont {Dotti}\ \emph
  {et~al.}(2007{\natexlab{a}})\citenamefont {Dotti}, \citenamefont {Oliva},\
  and\ \citenamefont {Troncoso}}]{Dotti:2006cp}%
  \BibitemOpen
  \bibfield  {author} {\bibinfo {author} {\bibfnamefont {G.}~\bibnamefont
  {Dotti}}, \bibinfo {author} {\bibfnamefont {J.}~\bibnamefont {Oliva}}, \ and\
  \bibinfo {author} {\bibfnamefont {R.}~\bibnamefont {Troncoso}},\ }\href
  {\doibase 10.1103/PhysRevD.75.024002} {\bibfield  {journal} {\bibinfo
  {journal} {Phys. Rev.}\ }\textbf {\bibinfo {volume} {D75}},\ \bibinfo {pages}
  {024002} (\bibinfo {year} {2007}{\natexlab{a}})}\BibitemShut {NoStop}%
\bibitem [{\citenamefont {Dotti}\ \emph
  {et~al.}(2007{\natexlab{b}})\citenamefont {Dotti}, \citenamefont {Oliva},\
  and\ \citenamefont {Troncoso}}]{Dotti:2007az}%
  \BibitemOpen
  \bibfield  {author} {\bibinfo {author} {\bibfnamefont {G.}~\bibnamefont
  {Dotti}}, \bibinfo {author} {\bibfnamefont {J.}~\bibnamefont {Oliva}}, \ and\
  \bibinfo {author} {\bibfnamefont {R.}~\bibnamefont {Troncoso}},\ }\href
  {\doibase 10.1103/PhysRevD.76.064038} {\bibfield  {journal} {\bibinfo
  {journal} {Phys. Rev.}\ }\textbf {\bibinfo {volume} {D76}},\ \bibinfo {pages}
  {064038} (\bibinfo {year} {2007}{\natexlab{b}})}\BibitemShut {NoStop}%
\bibitem [{\citenamefont {Cai}\ and\ \citenamefont {Ohta}(2006)}]{Cai:2006pq}%
  \BibitemOpen
  \bibfield  {author} {\bibinfo {author} {\bibfnamefont {R.-G.}\ \bibnamefont
  {Cai}}\ and\ \bibinfo {author} {\bibfnamefont {N.}~\bibnamefont {Ohta}},\
  }\href {\doibase 10.1103/PhysRevD.74.064001} {\bibfield  {journal} {\bibinfo
  {journal} {Phys. Rev.}\ }\textbf {\bibinfo {volume} {D74}},\ \bibinfo {pages}
  {064001} (\bibinfo {year} {2006})}\BibitemShut {NoStop}%
\bibitem [{\citenamefont {Garraffo}\ and\ \citenamefont
  {Giribet}(2008)}]{Garraffo:2008hu}%
  \BibitemOpen
  \bibfield  {author} {\bibinfo {author} {\bibfnamefont {C.}~\bibnamefont
  {Garraffo}}\ and\ \bibinfo {author} {\bibfnamefont {G.}~\bibnamefont
  {Giribet}},\ }\href {\doibase 10.1142/S0217732308027497} {\bibfield
  {journal} {\bibinfo  {journal} {Mod. Phys. Lett.}\ }\textbf {\bibinfo
  {volume} {A23}},\ \bibinfo {pages} {1801} (\bibinfo {year}
  {2008})}\BibitemShut {NoStop}%
\bibitem [{\citenamefont {Dehghani}\ \emph {et~al.}(2008)\citenamefont
  {Dehghani}, \citenamefont {Alinejadi},\ and\ \citenamefont
  {Hendi}}]{Dehghani:2008qr}%
  \BibitemOpen
  \bibfield  {author} {\bibinfo {author} {\bibfnamefont {M.~H.}\ \bibnamefont
  {Dehghani}}, \bibinfo {author} {\bibfnamefont {N.}~\bibnamefont {Alinejadi}},
  \ and\ \bibinfo {author} {\bibfnamefont {S.~H.}\ \bibnamefont {Hendi}},\
  }\href {\doibase 10.1103/PhysRevD.77.104025} {\bibfield  {journal} {\bibinfo
  {journal} {Phys. Rev.}\ }\textbf {\bibinfo {volume} {D77}},\ \bibinfo {pages}
  {104025} (\bibinfo {year} {2008})}\BibitemShut {NoStop}%
\bibitem [{\citenamefont {Anabalon}\ \emph {et~al.}(2009)\citenamefont
  {Anabalon}, \citenamefont {Deruelle}, \citenamefont {Morisawa}, \citenamefont
  {Oliva}, \citenamefont {Sasaki}, \citenamefont {Tempo},\ and\ \citenamefont
  {Troncoso}}]{Anabalon:2009kq}%
  \BibitemOpen
  \bibfield  {author} {\bibinfo {author} {\bibfnamefont {A.}~\bibnamefont
  {Anabalon}}, \bibinfo {author} {\bibfnamefont {N.}~\bibnamefont {Deruelle}},
  \bibinfo {author} {\bibfnamefont {Y.}~\bibnamefont {Morisawa}}, \bibinfo
  {author} {\bibfnamefont {J.}~\bibnamefont {Oliva}}, \bibinfo {author}
  {\bibfnamefont {M.}~\bibnamefont {Sasaki}}, \bibinfo {author} {\bibfnamefont
  {D.}~\bibnamefont {Tempo}}, \ and\ \bibinfo {author} {\bibfnamefont
  {R.}~\bibnamefont {Troncoso}},\ }\href {\doibase
  10.1088/0264-9381/26/6/065002} {\bibfield  {journal} {\bibinfo  {journal}
  {Class. Quant. Grav.}\ }\textbf {\bibinfo {volume} {26}},\ \bibinfo {pages}
  {065002} (\bibinfo {year} {2009})}\BibitemShut {NoStop}%
\bibitem [{\citenamefont {Dadhich}\ \emph {et~al.}(2013)\citenamefont
  {Dadhich}, \citenamefont {Pons},\ and\ \citenamefont
  {Prabhu}}]{Dadhich:2012ma}%
  \BibitemOpen
  \bibfield  {author} {\bibinfo {author} {\bibfnamefont {N.}~\bibnamefont
  {Dadhich}}, \bibinfo {author} {\bibfnamefont {J.~M.}\ \bibnamefont {Pons}}, \
  and\ \bibinfo {author} {\bibfnamefont {K.}~\bibnamefont {Prabhu}},\ }\href
  {\doibase 10.1007/s10714-013-1514-0} {\bibfield  {journal} {\bibinfo
  {journal} {Gen. Rel. Grav.}\ }\textbf {\bibinfo {volume} {45}},\ \bibinfo
  {pages} {1131} (\bibinfo {year} {2013})}\BibitemShut {NoStop}%
\bibitem [{\citenamefont {Aros}\ \emph {et~al.}(2000)\citenamefont {Aros},
  \citenamefont {Contreras}, \citenamefont {Olea}, \citenamefont {Troncoso},\
  and\ \citenamefont {Zanelli}}]{Aros:1999id}%
  \BibitemOpen
  \bibfield  {author} {\bibinfo {author} {\bibfnamefont {R.}~\bibnamefont
  {Aros}}, \bibinfo {author} {\bibfnamefont {M.}~\bibnamefont {Contreras}},
  \bibinfo {author} {\bibfnamefont {R.}~\bibnamefont {Olea}}, \bibinfo {author}
  {\bibfnamefont {R.}~\bibnamefont {Troncoso}}, \ and\ \bibinfo {author}
  {\bibfnamefont {J.}~\bibnamefont {Zanelli}},\ }\href {\doibase
  10.1103/PhysRevLett.84.1647} {\bibfield  {journal} {\bibinfo  {journal}
  {Phys. Rev. Lett.}\ }\textbf {\bibinfo {volume} {84}},\ \bibinfo {pages}
  {1647} (\bibinfo {year} {2000})}\BibitemShut {NoStop}%
\bibitem [{\citenamefont {Miskovic}\ and\ \citenamefont
  {Olea}(2009)}]{Miskovic:2009bm}%
  \BibitemOpen
  \bibfield  {author} {\bibinfo {author} {\bibfnamefont {O.}~\bibnamefont
  {Miskovic}}\ and\ \bibinfo {author} {\bibfnamefont {R.}~\bibnamefont
  {Olea}},\ }\href {\doibase 10.1103/PhysRevD.79.124020} {\bibfield  {journal}
  {\bibinfo  {journal} {Phys. Rev.}\ }\textbf {\bibinfo {volume} {D79}},\
  \bibinfo {pages} {124020} (\bibinfo {year} {2009})}\BibitemShut {NoStop}%
\bibitem [{\citenamefont {Camanho}\ \emph {et~al.}(2011)\citenamefont
  {Camanho}, \citenamefont {Edelstein},\ and\ \citenamefont
  {Paulos}}]{Camanho:2010ru}%
  \BibitemOpen
  \bibfield  {author} {\bibinfo {author} {\bibfnamefont {X.~O.}\ \bibnamefont
  {Camanho}}, \bibinfo {author} {\bibfnamefont {J.~D.}\ \bibnamefont
  {Edelstein}}, \ and\ \bibinfo {author} {\bibfnamefont {M.~F.}\ \bibnamefont
  {Paulos}},\ }\href {\doibase 10.1007/JHEP05(2011)127} {\bibfield  {journal}
  {\bibinfo  {journal} {JHEP}\ }\textbf {\bibinfo {volume} {05}},\ \bibinfo
  {pages} {127} (\bibinfo {year} {2011})}\BibitemShut {NoStop}%
\bibitem [{\citenamefont {Edelstein}(2014)}]{Edelstein:2013sna}%
  \BibitemOpen
  \bibfield  {author} {\bibinfo {author} {\bibfnamefont {J.~D.}\ \bibnamefont
  {Edelstein}},\ }\href {\doibase 10.1007/978-3-642-40157-2_2} {\bibfield
  {journal} {\bibinfo  {journal} {Springer Proc. Math. Stat.}\ }\textbf
  {\bibinfo {volume} {60}},\ \bibinfo {pages} {19} (\bibinfo {year}
  {2014})}\BibitemShut {NoStop}%
\bibitem [{\citenamefont {Cvetkovi\'c}\ \emph {et~al.}(2017)\citenamefont
  {Cvetkovi\'c}, \citenamefont {Miskovic},\ and\ \citenamefont
  {Simi\'c}}]{Cvetkovic:2017fxa}%
  \BibitemOpen
  \bibfield  {author} {\bibinfo {author} {\bibfnamefont {B.}~\bibnamefont
  {Cvetkovi\'c}}, \bibinfo {author} {\bibfnamefont {O.}~\bibnamefont
  {Miskovic}}, \ and\ \bibinfo {author} {\bibfnamefont {D.}~\bibnamefont
  {Simi\'c}},\ }\href {\doibase 10.1103/PhysRevD.96.044027} {\bibfield
  {journal} {\bibinfo  {journal} {Phys. Rev.}\ }\textbf {\bibinfo {volume}
  {D96}},\ \bibinfo {pages} {044027} (\bibinfo {year} {2017})}\BibitemShut
  {NoStop}%
\bibitem [{\citenamefont {Banados}\ \emph {et~al.}(2006)\citenamefont
  {Banados}, \citenamefont {Miskovic},\ and\ \citenamefont
  {Theisen}}]{Banados:2006fe}%
  \BibitemOpen
  \bibfield  {author} {\bibinfo {author} {\bibfnamefont {M.}~\bibnamefont
  {Banados}}, \bibinfo {author} {\bibfnamefont {O.}~\bibnamefont {Miskovic}}, \
  and\ \bibinfo {author} {\bibfnamefont {S.}~\bibnamefont {Theisen}},\ }\href
  {\doibase 10.1088/1126-6708/2006/06/025} {\bibfield  {journal} {\bibinfo
  {journal} {JHEP}\ }\textbf {\bibinfo {volume} {06}},\ \bibinfo {pages} {025}
  (\bibinfo {year} {2006})}\BibitemShut {NoStop}%
\bibitem [{\citenamefont {Gibbons}\ \emph {et~al.}(2003)\citenamefont
  {Gibbons}, \citenamefont {Ida},\ and\ \citenamefont
  {Shiromizu}}]{Gibbons:2002bh}%
  \BibitemOpen
  \bibfield  {author} {\bibinfo {author} {\bibfnamefont {G.~W.}\ \bibnamefont
  {Gibbons}}, \bibinfo {author} {\bibfnamefont {D.}~\bibnamefont {Ida}}, \ and\
  \bibinfo {author} {\bibfnamefont {T.}~\bibnamefont {Shiromizu}},\ }\href
  {\doibase 10.1143/PTPS.148.284} {\bibfield  {journal} {\bibinfo  {journal}
  {Prog. Theor. Phys. Suppl.}\ }\textbf {\bibinfo {volume} {148}},\ \bibinfo
  {pages} {284} (\bibinfo {year} {2003})}\BibitemShut {NoStop}%
\bibitem [{\citenamefont {Ray}(2015)}]{Ray:2015ava}%
  \BibitemOpen
  \bibfield  {author} {\bibinfo {author} {\bibfnamefont {S.}~\bibnamefont
  {Ray}},\ }\href {\doibase 10.1088/0264-9381/32/19/195022} {\bibfield
  {journal} {\bibinfo  {journal} {Class. Quant. Grav.}\ }\textbf {\bibinfo
  {volume} {32}},\ \bibinfo {pages} {195022} (\bibinfo {year}
  {2015})}\BibitemShut {NoStop}%
\bibitem [{\citenamefont {Wheeler}(1986)}]{Wheeler:1985qd}%
  \BibitemOpen
  \bibfield  {author} {\bibinfo {author} {\bibfnamefont {J.~T.}\ \bibnamefont
  {Wheeler}},\ }\href {\doibase 10.1016/0550-3213(86)90388-3} {\bibfield
  {journal} {\bibinfo  {journal} {Nucl. Phys.}\ }\textbf {\bibinfo {volume}
  {B273}},\ \bibinfo {pages} {732} (\bibinfo {year} {1986})}\BibitemShut
  {NoStop}%
\bibitem [{\citenamefont {Cai}(2004)}]{Cai:2003kt}%
  \BibitemOpen
  \bibfield  {author} {\bibinfo {author} {\bibfnamefont {R.-G.}\ \bibnamefont
  {Cai}},\ }\href {\doibase 10.1016/j.physletb.2004.01.015} {\bibfield
  {journal} {\bibinfo  {journal} {Phys. Lett.}\ }\textbf {\bibinfo {volume}
  {B582}},\ \bibinfo {pages} {237} (\bibinfo {year} {2004})}\BibitemShut
  {NoStop}%
\bibitem [{\citenamefont {Kastor}\ \emph {et~al.}(2011)\citenamefont {Kastor},
  \citenamefont {Ray},\ and\ \citenamefont {Traschen}}]{Kastor:2011qp}%
  \BibitemOpen
  \bibfield  {author} {\bibinfo {author} {\bibfnamefont {D.}~\bibnamefont
  {Kastor}}, \bibinfo {author} {\bibfnamefont {S.}~\bibnamefont {Ray}}, \ and\
  \bibinfo {author} {\bibfnamefont {J.}~\bibnamefont {Traschen}},\ }\href
  {\doibase 10.1088/0264-9381/28/19/195022} {\bibfield  {journal} {\bibinfo
  {journal} {Class. Quant. Grav.}\ }\textbf {\bibinfo {volume} {28}},\ \bibinfo
  {pages} {195022} (\bibinfo {year} {2011})}\BibitemShut {NoStop}%
\bibitem [{\citenamefont {Taub}(1951)}]{Taub:1950ez}%
  \BibitemOpen
  \bibfield  {author} {\bibinfo {author} {\bibfnamefont {A.~H.}\ \bibnamefont
  {Taub}},\ }\href {\doibase 10.2307/1969567} {\bibfield  {journal} {\bibinfo
  {journal} {Annals Math.}\ }\textbf {\bibinfo {volume} {53}},\ \bibinfo
  {pages} {472} (\bibinfo {year} {1951})}\BibitemShut {NoStop}%
\bibitem [{\citenamefont {Newman}\ \emph {et~al.}(1963)\citenamefont {Newman},
  \citenamefont {Tamubrino},\ and\ \citenamefont {Unti}}]{Newman:1963yy}%
  \BibitemOpen
  \bibfield  {author} {\bibinfo {author} {\bibfnamefont {E.}~\bibnamefont
  {Newman}}, \bibinfo {author} {\bibfnamefont {L.}~\bibnamefont {Tamubrino}}, \
  and\ \bibinfo {author} {\bibfnamefont {T.}~\bibnamefont {Unti}},\ }\href
  {\doibase 10.1063/1.1704018} {\bibfield  {journal} {\bibinfo  {journal} {J.
  Math. Phys.}\ }\textbf {\bibinfo {volume} {4}},\ \bibinfo {pages} {915}
  (\bibinfo {year} {1963})}\BibitemShut {NoStop}%
\bibitem [{\citenamefont {Chakraborty}\ and\ \citenamefont
  {Bhattacharyya}(2018)}]{Chakraborty:2017nfu}%
  \BibitemOpen
  \bibfield  {author} {\bibinfo {author} {\bibfnamefont {C.}~\bibnamefont
  {Chakraborty}}\ and\ \bibinfo {author} {\bibfnamefont {S.}~\bibnamefont
  {Bhattacharyya}},\ }\href {\doibase 10.1103/PhysRevD.98.043021} {\bibfield
  {journal} {\bibinfo  {journal} {Phys. Rev.}\ }\textbf {\bibinfo {volume}
  {D98}},\ \bibinfo {pages} {043021} (\bibinfo {year} {2018})}\BibitemShut
  {NoStop}%
\bibitem [{\citenamefont {Chakraborty}\ and\ \citenamefont
  {Bhattacharyya}(2019)}]{Chakraborty:2019rna}%
  \BibitemOpen
  \bibfield  {author} {\bibinfo {author} {\bibfnamefont {C.}~\bibnamefont
  {Chakraborty}}\ and\ \bibinfo {author} {\bibfnamefont {S.}~\bibnamefont
  {Bhattacharyya}},\ }\href {\doibase 10.1088/1475-7516/2019/05/034} {\bibfield
   {journal} {\bibinfo  {journal} {JCAP}\ }\textbf {\bibinfo {volume} {1905}},\
  \bibinfo {pages} {034} (\bibinfo {year} {2019})}\BibitemShut {NoStop}%
\bibitem [{\citenamefont {Bais}\ and\ \citenamefont
  {Batenburg}(1985)}]{Bais:1984xb}%
  \BibitemOpen
  \bibfield  {author} {\bibinfo {author} {\bibfnamefont {F.~A.}\ \bibnamefont
  {Bais}}\ and\ \bibinfo {author} {\bibfnamefont {P.}~\bibnamefont
  {Batenburg}},\ }\href {\doibase 10.1016/0550-3213(85)90524-3} {\bibfield
  {journal} {\bibinfo  {journal} {Nucl. Phys.}\ }\textbf {\bibinfo {volume}
  {B253}},\ \bibinfo {pages} {162} (\bibinfo {year} {1985})}\BibitemShut
  {NoStop}%
\bibitem [{\citenamefont {Page}\ and\ \citenamefont
  {Pope}(1987)}]{Page:1985bq}%
  \BibitemOpen
  \bibfield  {author} {\bibinfo {author} {\bibfnamefont {D.~N.}\ \bibnamefont
  {Page}}\ and\ \bibinfo {author} {\bibfnamefont {C.~N.}\ \bibnamefont
  {Pope}},\ }\href {\doibase 10.1088/0264-9381/4/2/005} {\bibfield  {journal}
  {\bibinfo  {journal} {Class. Quant. Grav.}\ }\textbf {\bibinfo {volume}
  {4}},\ \bibinfo {pages} {213} (\bibinfo {year} {1987})}\BibitemShut {NoStop}%
\bibitem [{\citenamefont {Taylor}()}]{Taylor:1998fd}%
  \BibitemOpen
  \bibfield  {author} {\bibinfo {author} {\bibfnamefont {M.}~\bibnamefont
  {Taylor}},\ }\href@noop {} {\ }\Eprint {http://arxiv.org/abs/hep-th/9809041}
  {hep-th/9809041} \BibitemShut {NoStop}%
\bibitem [{\citenamefont {Flores-Alfonso}\ and\ \citenamefont
  {Quevedo}(2019)}]{Flores-Alfonso:2018jra}%
  \BibitemOpen
  \bibfield  {author} {\bibinfo {author} {\bibfnamefont {D.}~\bibnamefont
  {Flores-Alfonso}}\ and\ \bibinfo {author} {\bibfnamefont {H.}~\bibnamefont
  {Quevedo}},\ }\href {\doibase 10.1142/S0219887819501548} {\bibfield
  {journal} {\bibinfo  {journal} {Int. J. Geom. Methods Mod. Phys.}\ }\textbf
  {\bibinfo {volume} {16}},\ \bibinfo {pages} {1950154} (\bibinfo {year}
  {2019})}\BibitemShut {NoStop}%
\bibitem [{\citenamefont {Sorkin}(1983)}]{Sorkin:1983ns}%
  \BibitemOpen
  \bibfield  {author} {\bibinfo {author} {\bibfnamefont {R.~d.}\ \bibnamefont
  {Sorkin}},\ }\href {\doibase 10.1103/PhysRevLett.51.87} {\bibfield  {journal}
  {\bibinfo  {journal} {Phys. Rev. Lett.}\ }\textbf {\bibinfo {volume} {51}},\
  \bibinfo {pages} {87} (\bibinfo {year} {1983})}\BibitemShut {NoStop}%
\bibitem [{\citenamefont {Hashemi}\ and\ \citenamefont
  {Riazi}(2018)}]{Hashemi:2018jbv}%
  \BibitemOpen
  \bibfield  {author} {\bibinfo {author} {\bibfnamefont {S.~S.}\ \bibnamefont
  {Hashemi}}\ and\ \bibinfo {author} {\bibfnamefont {N.}~\bibnamefont
  {Riazi}},\ }\href {\doibase 10.1016/j.aop.2018.04.004} {\bibfield  {journal}
  {\bibinfo  {journal} {Annals Phys.}\ }\textbf {\bibinfo {volume} {393}},\
  \bibinfo {pages} {206} (\bibinfo {year} {2018})}\BibitemShut {NoStop}%
\bibitem [{\citenamefont {Sedigheh~Hashemi}\ and\ \citenamefont
  {Riazi}(2018)}]{Hashemi:2018ujp}%
  \BibitemOpen
  \bibfield  {author} {\bibinfo {author} {\bibfnamefont {S.}~\bibnamefont
  {Sedigheh~Hashemi}}\ and\ \bibinfo {author} {\bibfnamefont {N.}~\bibnamefont
  {Riazi}},\ }\href {\doibase 10.1007/s10714-018-2343-y} {\bibfield  {journal}
  {\bibinfo  {journal} {Gen. Rel. Grav.}\ }\textbf {\bibinfo {volume} {50}},\
  \bibinfo {pages} {19} (\bibinfo {year} {2018})}\BibitemShut {NoStop}%
\bibitem [{\citenamefont {Linshaw}\ and\ \citenamefont
  {Mathai}(2018)}]{Linshaw:2017bpf}%
  \BibitemOpen
  \bibfield  {author} {\bibinfo {author} {\bibfnamefont {A.}~\bibnamefont
  {Linshaw}}\ and\ \bibinfo {author} {\bibfnamefont {V.}~\bibnamefont
  {Mathai}},\ }\href {\doibase 10.1016/j.geomphys.2018.03.017} {\bibfield
  {journal} {\bibinfo  {journal} {J. Geom. Phys.}\ }\textbf {\bibinfo {volume}
  {129}},\ \bibinfo {pages} {269} (\bibinfo {year} {2018})}\BibitemShut
  {NoStop}%
\bibitem [{\citenamefont {Li}(2019)}]{Li2019}%
  \BibitemOpen
  \bibfield  {author} {\bibinfo {author} {\bibfnamefont {Y.}~\bibnamefont
  {Li}},\ }\href {\doibase 10.1007/s00222-019-00861-w} {\bibfield  {journal}
  {\bibinfo  {journal} {Inventiones mathematicae}\ } (\bibinfo {year} {2019}),\
  10.1007/s00222-019-00861-w}\BibitemShut {NoStop}%
\bibitem [{\citenamefont {Foscolo}\ \emph {et~al.}()\citenamefont {Foscolo},
  \citenamefont {Haskins},\ and\ \citenamefont
  {Nordstr{\"{o}}m}}]{Foscolo:2018mfs}%
  \BibitemOpen
  \bibfield  {author} {\bibinfo {author} {\bibfnamefont {L.}~\bibnamefont
  {Foscolo}}, \bibinfo {author} {\bibfnamefont {M.}~\bibnamefont {Haskins}}, \
  and\ \bibinfo {author} {\bibfnamefont {J.}~\bibnamefont {Nordstr{\"{o}}m}},\
  }\href@noop {} {\ }\Eprint {http://arxiv.org/abs/arXiv:1805.02612}
  {arXiv:1805.02612} \BibitemShut {NoStop}%
\bibitem [{\citenamefont {Johnson}(2018)}]{Johnson:2017ood}%
  \BibitemOpen
  \bibfield  {author} {\bibinfo {author} {\bibfnamefont {C.~V.}\ \bibnamefont
  {Johnson}},\ }\href {\doibase 10.1088/1361-6382/aaa010} {\bibfield  {journal}
  {\bibinfo  {journal} {Class. Quant. Grav.}\ }\textbf {\bibinfo {volume}
  {35}},\ \bibinfo {pages} {045001} (\bibinfo {year} {2018})}\BibitemShut
  {NoStop}%
\bibitem [{\citenamefont {Flores-Alfonso}\ and\ \citenamefont
  {Quevedo}(2017)}]{Flores-Alfonso:2017kvy}%
  \BibitemOpen
  \bibfield  {author} {\bibinfo {author} {\bibfnamefont {D.}~\bibnamefont
  {Flores-Alfonso}}\ and\ \bibinfo {author} {\bibfnamefont {H.}~\bibnamefont
  {Quevedo}},\ }\href {\doibase 10.7546/jgsp-44-2017-39-54} {\bibfield
  {journal} {\bibinfo  {journal} {J. Geom. Symmmetry Phys.}\ }\textbf {\bibinfo
  {volume} {44}},\ \bibinfo {pages} {39} (\bibinfo {year} {2017})}\BibitemShut
  {NoStop}%
\bibitem [{\citenamefont {Brill}(1964)}]{Brill1964}%
  \BibitemOpen
  \bibfield  {author} {\bibinfo {author} {\bibfnamefont {D.~R.}\ \bibnamefont
  {Brill}},\ }\href {\doibase 10.1103/PhysRev.133.B845} {\bibfield  {journal}
  {\bibinfo  {journal} {Phys. Rev.}\ }\textbf {\bibinfo {volume} {133}},\
  \bibinfo {pages} {B845} (\bibinfo {year} {1964})}\BibitemShut {NoStop}%
\bibitem [{\citenamefont {Carter}(1968)}]{Carter:1968ks}%
  \BibitemOpen
  \bibfield  {author} {\bibinfo {author} {\bibfnamefont {B.}~\bibnamefont
  {Carter}},\ }\href {\doibase 10.1007/BF03399503} {\bibfield  {journal}
  {\bibinfo  {journal} {Commun. Math. Phys.}\ }\textbf {\bibinfo {volume}
  {10}},\ \bibinfo {pages} {280} (\bibinfo {year} {1968})}\BibitemShut
  {NoStop}%
\bibitem [{\citenamefont {Page}(1978)}]{Page:1979aj}%
  \BibitemOpen
  \bibfield  {author} {\bibinfo {author} {\bibfnamefont {D.~N.}\ \bibnamefont
  {Page}},\ }\href {\doibase 10.1016/0370-2693(78)90016-3} {\bibfield
  {journal} {\bibinfo  {journal} {Phys. Lett.}\ }\textbf {\bibinfo {volume}
  {78B}},\ \bibinfo {pages} {249} (\bibinfo {year} {1978})}\BibitemShut
  {NoStop}%
\bibitem [{\citenamefont {Garc{\'i}a~D.}\ \emph {et~al.}(1984)\citenamefont
  {Garc{\'i}a~D.}, \citenamefont {Salazar~I.},\ and\ \citenamefont
  {Pleba{\'{n}}ski}}]{GarciaD.1984}%
  \BibitemOpen
  \bibfield  {author} {\bibinfo {author} {\bibfnamefont {A.}~\bibnamefont
  {Garc{\'i}a~D.}}, \bibinfo {author} {\bibfnamefont {H.}~\bibnamefont
  {Salazar~I.}}, \ and\ \bibinfo {author} {\bibfnamefont {J.~F.}\ \bibnamefont
  {Pleba{\'{n}}ski}},\ }\href {\doibase 10.1007/BF02721649} {\bibfield
  {journal} {\bibinfo  {journal} {Il Nuovo Cimento B (1971-1996)}\ }\textbf
  {\bibinfo {volume} {84}},\ \bibinfo {pages} {65} (\bibinfo {year}
  {1984})}\BibitemShut {NoStop}%
\bibitem [{\citenamefont {Awad}\ and\ \citenamefont
  {Chamblin}(2002)}]{Awad:2000gg}%
  \BibitemOpen
  \bibfield  {author} {\bibinfo {author} {\bibfnamefont {A.}~\bibnamefont
  {Awad}}\ and\ \bibinfo {author} {\bibfnamefont {A.}~\bibnamefont
  {Chamblin}},\ }\href {\doibase 10.1088/0264-9381/19/8/301} {\bibfield
  {journal} {\bibinfo  {journal} {Class. Quant. Grav.}\ }\textbf {\bibinfo
  {volume} {19}},\ \bibinfo {pages} {2051} (\bibinfo {year}
  {2002})}\BibitemShut {NoStop}%
\bibitem [{\citenamefont {Awad}(2006)}]{Awad:2005ff}%
  \BibitemOpen
  \bibfield  {author} {\bibinfo {author} {\bibfnamefont {A.~M.}\ \bibnamefont
  {Awad}},\ }\href {\doibase 10.1088/0264-9381/23/9/006} {\bibfield  {journal}
  {\bibinfo  {journal} {Class. Quant. Grav.}\ }\textbf {\bibinfo {volume}
  {23}},\ \bibinfo {pages} {2849} (\bibinfo {year} {2006})}\BibitemShut
  {NoStop}%
\bibitem [{\citenamefont {Bueno}\ \emph {et~al.}(2018)\citenamefont {Bueno},
  \citenamefont {Cano}, \citenamefont {Hennigar},\ and\ \citenamefont
  {Mann}}]{Bueno:2018uoy}%
  \BibitemOpen
  \bibfield  {author} {\bibinfo {author} {\bibfnamefont {P.}~\bibnamefont
  {Bueno}}, \bibinfo {author} {\bibfnamefont {P.~A.}\ \bibnamefont {Cano}},
  \bibinfo {author} {\bibfnamefont {R.~A.}\ \bibnamefont {Hennigar}}, \ and\
  \bibinfo {author} {\bibfnamefont {R.~B.}\ \bibnamefont {Mann}},\ }\href
  {\doibase 10.1007/JHEP10(2018)095} {\bibfield  {journal} {\bibinfo  {journal}
  {JHEP}\ }\textbf {\bibinfo {volume} {10}},\ \bibinfo {pages} {095} (\bibinfo
  {year} {2018})}\BibitemShut {NoStop}%
\bibitem [{\citenamefont {Hendi}\ and\ \citenamefont
  {Dehghani}(2008)}]{Hendi:2008wq}%
  \BibitemOpen
  \bibfield  {author} {\bibinfo {author} {\bibfnamefont {S.~H.}\ \bibnamefont
  {Hendi}}\ and\ \bibinfo {author} {\bibfnamefont {M.~H.}\ \bibnamefont
  {Dehghani}},\ }\href {\doibase 10.1016/j.physletb.2008.07.002} {\bibfield
  {journal} {\bibinfo  {journal} {Phys. Lett.}\ }\textbf {\bibinfo {volume}
  {B666}},\ \bibinfo {pages} {116} (\bibinfo {year} {2008})}\BibitemShut
  {NoStop}%
\bibitem [{\citenamefont {Dehghani}\ and\ \citenamefont
  {Hendi}(2006)}]{Dehghani:2006aa}%
  \BibitemOpen
  \bibfield  {author} {\bibinfo {author} {\bibfnamefont {M.~H.}\ \bibnamefont
  {Dehghani}}\ and\ \bibinfo {author} {\bibfnamefont {S.~H.}\ \bibnamefont
  {Hendi}},\ }\href {\doibase 10.1103/PhysRevD.73.084021} {\bibfield  {journal}
  {\bibinfo  {journal} {Phys. Rev.}\ }\textbf {\bibinfo {volume} {D73}},\
  \bibinfo {pages} {084021} (\bibinfo {year} {2006})}\BibitemShut {NoStop}%
\bibitem [{\citenamefont {Mardones}\ and\ \citenamefont
  {Zanelli}(1991)}]{Mardones:1990qc}%
  \BibitemOpen
  \bibfield  {author} {\bibinfo {author} {\bibfnamefont {A.}~\bibnamefont
  {Mardones}}\ and\ \bibinfo {author} {\bibfnamefont {J.}~\bibnamefont
  {Zanelli}},\ }\href {\doibase 10.1088/0264-9381/8/8/018} {\bibfield
  {journal} {\bibinfo  {journal} {Class. Quant. Grav.}\ }\textbf {\bibinfo
  {volume} {8}},\ \bibinfo {pages} {1545} (\bibinfo {year} {1991})}\BibitemShut
  {NoStop}%
\bibitem [{\citenamefont {Hehl}\ \emph {et~al.}(1976)\citenamefont {Hehl},
  \citenamefont {Von Der~Heyde}, \citenamefont {Kerlick},\ and\ \citenamefont
  {Nester}}]{Hehl:1976kj}%
  \BibitemOpen
  \bibfield  {author} {\bibinfo {author} {\bibfnamefont {F.~W.}\ \bibnamefont
  {Hehl}}, \bibinfo {author} {\bibfnamefont {P.}~\bibnamefont {Von Der~Heyde}},
  \bibinfo {author} {\bibfnamefont {G.~D.}\ \bibnamefont {Kerlick}}, \ and\
  \bibinfo {author} {\bibfnamefont {J.~M.}\ \bibnamefont {Nester}},\ }\href
  {\doibase 10.1103/RevModPhys.48.393} {\bibfield  {journal} {\bibinfo
  {journal} {Rev. Mod. Phys.}\ }\textbf {\bibinfo {volume} {48}},\ \bibinfo
  {pages} {393} (\bibinfo {year} {1976})}\BibitemShut {NoStop}%
\bibitem [{\citenamefont {Hehl}\ \emph {et~al.}(1995)\citenamefont {Hehl},
  \citenamefont {McCrea}, \citenamefont {Mielke},\ and\ \citenamefont
  {Ne'eman}}]{Hehl:1994ue}%
  \BibitemOpen
  \bibfield  {author} {\bibinfo {author} {\bibfnamefont {F.~W.}\ \bibnamefont
  {Hehl}}, \bibinfo {author} {\bibfnamefont {J.~D.}\ \bibnamefont {McCrea}},
  \bibinfo {author} {\bibfnamefont {E.~W.}\ \bibnamefont {Mielke}}, \ and\
  \bibinfo {author} {\bibfnamefont {Y.}~\bibnamefont {Ne'eman}},\ }\href
  {\doibase 10.1016/0370-1573(94)00111-F} {\bibfield  {journal} {\bibinfo
  {journal} {Phys. Rept.}\ }\textbf {\bibinfo {volume} {258}},\ \bibinfo
  {pages} {1} (\bibinfo {year} {1995})}\BibitemShut {NoStop}%
\bibitem [{\citenamefont {Corral}\ and\ \citenamefont
  {Bonder}(2019)}]{Corral:2018hxi}%
  \BibitemOpen
  \bibfield  {author} {\bibinfo {author} {\bibfnamefont {C.}~\bibnamefont
  {Corral}}\ and\ \bibinfo {author} {\bibfnamefont {Y.}~\bibnamefont
  {Bonder}},\ }\href {\doibase 10.1088/1361-6382/aafce1} {\bibfield  {journal}
  {\bibinfo  {journal} {{Class. Quant. Grav.}}\ }\textbf {\bibinfo {volume}
  {36}},\ \bibinfo {pages} {045002} (\bibinfo {year} {2019})}\BibitemShut
  {NoStop}%
\bibitem [{\citenamefont {Troncoso}\ and\ \citenamefont
  {Zanelli}(2000)}]{Troncoso:1999pk}%
  \BibitemOpen
  \bibfield  {author} {\bibinfo {author} {\bibfnamefont {R.}~\bibnamefont
  {Troncoso}}\ and\ \bibinfo {author} {\bibfnamefont {J.}~\bibnamefont
  {Zanelli}},\ }\href {\doibase 10.1088/0264-9381/17/21/307} {\bibfield
  {journal} {\bibinfo  {journal} {Class. Quant. Grav.}\ }\textbf {\bibinfo
  {volume} {17}},\ \bibinfo {pages} {4451} (\bibinfo {year}
  {2000})}\BibitemShut {NoStop}%
\bibitem [{\citenamefont {Canfora}\ \emph {et~al.}(2007)\citenamefont
  {Canfora}, \citenamefont {Giacomini},\ and\ \citenamefont
  {Willison}}]{Canfora:2007ux}%
  \BibitemOpen
  \bibfield  {author} {\bibinfo {author} {\bibfnamefont {F.}~\bibnamefont
  {Canfora}}, \bibinfo {author} {\bibfnamefont {A.}~\bibnamefont {Giacomini}},
  \ and\ \bibinfo {author} {\bibfnamefont {S.}~\bibnamefont {Willison}},\
  }\href {\doibase 10.1103/PhysRevD.76.044021} {\bibfield  {journal} {\bibinfo
  {journal} {Phys. Rev.}\ }\textbf {\bibinfo {volume} {D76}},\ \bibinfo {pages}
  {044021} (\bibinfo {year} {2007})}\BibitemShut {NoStop}%
\bibitem [{\citenamefont {Canfora}\ \emph {et~al.}(2008)\citenamefont
  {Canfora}, \citenamefont {Giacomini},\ and\ \citenamefont
  {Troncoso}}]{Canfora:2007xs}%
  \BibitemOpen
  \bibfield  {author} {\bibinfo {author} {\bibfnamefont {F.}~\bibnamefont
  {Canfora}}, \bibinfo {author} {\bibfnamefont {A.}~\bibnamefont {Giacomini}},
  \ and\ \bibinfo {author} {\bibfnamefont {R.}~\bibnamefont {Troncoso}},\
  }\href {\doibase 10.1103/PhysRevD.77.024002} {\bibfield  {journal} {\bibinfo
  {journal} {Phys. Rev.}\ }\textbf {\bibinfo {volume} {D77}},\ \bibinfo {pages}
  {024002} (\bibinfo {year} {2008})}\BibitemShut {NoStop}%
\bibitem [{\citenamefont {Canfora}\ and\ \citenamefont
  {Giacomini}(2008)}]{Canfora:2008ka}%
  \BibitemOpen
  \bibfield  {author} {\bibinfo {author} {\bibfnamefont {F.}~\bibnamefont
  {Canfora}}\ and\ \bibinfo {author} {\bibfnamefont {A.}~\bibnamefont
  {Giacomini}},\ }\href {\doibase 10.1103/PhysRevD.78.084034} {\bibfield
  {journal} {\bibinfo  {journal} {Phys. Rev.}\ }\textbf {\bibinfo {volume}
  {D78}},\ \bibinfo {pages} {084034} (\bibinfo {year} {2008})}\BibitemShut
  {NoStop}%
\bibitem [{\citenamefont {Canfora}\ and\ \citenamefont
  {Giacomini}(2010)}]{Canfora:2010rh}%
  \BibitemOpen
  \bibfield  {author} {\bibinfo {author} {\bibfnamefont {F.}~\bibnamefont
  {Canfora}}\ and\ \bibinfo {author} {\bibfnamefont {A.}~\bibnamefont
  {Giacomini}},\ }\href {\doibase 10.1103/PhysRevD.82.024022} {\bibfield
  {journal} {\bibinfo  {journal} {Phys. Rev.}\ }\textbf {\bibinfo {volume}
  {D82}},\ \bibinfo {pages} {024022} (\bibinfo {year} {2010})}\BibitemShut
  {NoStop}%
\bibitem [{\citenamefont {Cvetkovi\'c}\ and\ \citenamefont
  {Simi\'c}(2016)}]{Cvetkovic:2016ios}%
  \BibitemOpen
  \bibfield  {author} {\bibinfo {author} {\bibfnamefont {B.}~\bibnamefont
  {Cvetkovi\'c}}\ and\ \bibinfo {author} {\bibfnamefont {D.}~\bibnamefont
  {Simi\'c}},\ }\href {\doibase 10.1103/PhysRevD.94.084037} {\bibfield
  {journal} {\bibinfo  {journal} {Phys. Rev.}\ }\textbf {\bibinfo {volume}
  {D94}},\ \bibinfo {pages} {084037} (\bibinfo {year} {2016})}\BibitemShut
  {NoStop}%
\bibitem [{\citenamefont {Cvetkovi\'c}\ and\ \citenamefont
  {Simi\'c}(2018)}]{Cvetkovic:2017nkg}%
  \BibitemOpen
  \bibfield  {author} {\bibinfo {author} {\bibfnamefont {B.}~\bibnamefont
  {Cvetkovi\'c}}\ and\ \bibinfo {author} {\bibfnamefont {D.}~\bibnamefont
  {Simi\'c}},\ }\href {\doibase 10.1088/1361-6382/aaa3a7} {\bibfield  {journal}
  {\bibinfo  {journal} {Class. Quant. Grav.}\ }\textbf {\bibinfo {volume}
  {35}},\ \bibinfo {pages} {055005} (\bibinfo {year} {2018})}\BibitemShut
  {NoStop}%
\bibitem [{\citenamefont {Dehghani}\ and\ \citenamefont
  {Mann}(2005)}]{Dehghani:2005zm}%
  \BibitemOpen
  \bibfield  {author} {\bibinfo {author} {\bibfnamefont {M.~H.}\ \bibnamefont
  {Dehghani}}\ and\ \bibinfo {author} {\bibfnamefont {R.~B.}\ \bibnamefont
  {Mann}},\ }\href {\doibase 10.1103/PhysRevD.72.124006} {\bibfield  {journal}
  {\bibinfo  {journal} {Phys. Rev.}\ }\textbf {\bibinfo {volume} {D72}},\
  \bibinfo {pages} {124006} (\bibinfo {year} {2005})}\BibitemShut {NoStop}%
\bibitem [{\citenamefont {Araneda}\ \emph {et~al.}(2016)\citenamefont
  {Araneda}, \citenamefont {Aros}, \citenamefont {Miskovic},\ and\
  \citenamefont {Olea}}]{Araneda:2016iiy}%
  \BibitemOpen
  \bibfield  {author} {\bibinfo {author} {\bibfnamefont {R.}~\bibnamefont
  {Araneda}}, \bibinfo {author} {\bibfnamefont {R.}~\bibnamefont {Aros}},
  \bibinfo {author} {\bibfnamefont {O.}~\bibnamefont {Miskovic}}, \ and\
  \bibinfo {author} {\bibfnamefont {R.}~\bibnamefont {Olea}},\ }\href {\doibase
  10.1103/PhysRevD.93.084022} {\bibfield  {journal} {\bibinfo  {journal} {Phys.
  Rev.}\ }\textbf {\bibinfo {volume} {D93}},\ \bibinfo {pages} {084022}
  (\bibinfo {year} {2016})}\BibitemShut {NoStop}%
\bibitem [{\citenamefont {Bordo}\ \emph {et~al.}()\citenamefont {Bordo},
  \citenamefont {Gray}, \citenamefont {Hennigar},\ and\ \citenamefont
  {Kubiz{\v{n}}{\'{a}}k}}]{Bordo:2019tyh}%
  \BibitemOpen
  \bibfield  {author} {\bibinfo {author} {\bibfnamefont {A.~B.}\ \bibnamefont
  {Bordo}}, \bibinfo {author} {\bibfnamefont {F.}~\bibnamefont {Gray}},
  \bibinfo {author} {\bibfnamefont {R.~A.}\ \bibnamefont {Hennigar}}, \ and\
  \bibinfo {author} {\bibfnamefont {D.}~\bibnamefont {Kubiz{\v{n}}{\'{a}}k}},\
  }\href@noop {} {\ }\Eprint {http://arxiv.org/abs/arXiv:1905.03785}
  {arXiv:1905.03785} \BibitemShut {NoStop}%
\bibitem [{\citenamefont {Bishop}\ and\ \citenamefont
  {Goldberg}(1965)}]{Bishop:1965}%
  \BibitemOpen
  \bibfield  {author} {\bibinfo {author} {\bibfnamefont {R.}~\bibnamefont
  {Bishop}}\ and\ \bibinfo {author} {\bibfnamefont {S.}~\bibnamefont
  {Goldberg}},\ }\href {\doibase
  https://doi.org/10.1090/S0002-9939-1965-0172221-6} {\bibfield  {journal}
  {\bibinfo  {journal} {Proc. Amer. Math. Soc.}\ }\textbf {\bibinfo {volume}
  {16}},\ \bibinfo {pages} {119} (\bibinfo {year} {1965})}\BibitemShut
  {NoStop}%
\bibitem [{\citenamefont {Hawking}\ and\ \citenamefont
  {Page}(1983)}]{Hawking:1982dh}%
  \BibitemOpen
  \bibfield  {author} {\bibinfo {author} {\bibfnamefont {S.~W.}\ \bibnamefont
  {Hawking}}\ and\ \bibinfo {author} {\bibfnamefont {D.~N.}\ \bibnamefont
  {Page}},\ }\href {\doibase 10.1007/BF01208266} {\bibfield  {journal}
  {\bibinfo  {journal} {Commun. Math. Phys.}\ }\textbf {\bibinfo {volume}
  {87}},\ \bibinfo {pages} {577} (\bibinfo {year} {1983})}\BibitemShut
  {NoStop}%
\bibitem [{\citenamefont {Johnson}(2014)}]{Johnson:2014pwa}%
  \BibitemOpen
  \bibfield  {author} {\bibinfo {author} {\bibfnamefont {C.~V.}\ \bibnamefont
  {Johnson}},\ }\href {\doibase 10.1088/0264-9381/31/22/225005} {\bibfield
  {journal} {\bibinfo  {journal} {Class. Quant. Grav.}\ }\textbf {\bibinfo
  {volume} {31}},\ \bibinfo {pages} {225005} (\bibinfo {year}
  {2014})}\BibitemShut {NoStop}%
\bibitem [{\citenamefont {Hoxha}\ \emph {et~al.}(2000)\citenamefont {Hoxha},
  \citenamefont {Martinez-Acosta},\ and\ \citenamefont {Pope}}]{Hoxha:2000jf}%
  \BibitemOpen
  \bibfield  {author} {\bibinfo {author} {\bibfnamefont {P.}~\bibnamefont
  {Hoxha}}, \bibinfo {author} {\bibfnamefont {R.~R.}\ \bibnamefont
  {Martinez-Acosta}}, \ and\ \bibinfo {author} {\bibfnamefont {C.~N.}\
  \bibnamefont {Pope}},\ }\href {\doibase 10.1088/0264-9381/17/20/305}
  {\bibfield  {journal} {\bibinfo  {journal} {Class. Quant. Grav.}\ }\textbf
  {\bibinfo {volume} {17}},\ \bibinfo {pages} {4207} (\bibinfo {year}
  {2000})}\BibitemShut {NoStop}%
\bibitem [{\citenamefont {Hennigar}\ \emph {et~al.}()\citenamefont {Hennigar},
  \citenamefont {Kubiz{\v{n}}{\'{a}}k},\ and\ \citenamefont
  {Mann}}]{Kubiznak:2019yiu}%
  \BibitemOpen
  \bibfield  {author} {\bibinfo {author} {\bibfnamefont {R.~A.}\ \bibnamefont
  {Hennigar}}, \bibinfo {author} {\bibfnamefont {D.}~\bibnamefont
  {Kubiz{\v{n}}{\'{a}}k}}, \ and\ \bibinfo {author} {\bibfnamefont {R.~B.}\
  \bibnamefont {Mann}},\ }\href@noop {} {\ }\Eprint
  {http://arxiv.org/abs/arXiv:1903.08668} {arXiv:1903.08668} \BibitemShut
  {NoStop}%
\bibitem [{\citenamefont {Bordo}\ \emph {et~al.}(2019)\citenamefont {Bordo},
  \citenamefont {Gray},\ and\ \citenamefont
  {Kubiz{\v{n}}{\'{a}}k}}]{Ballon:2019uha}%
  \BibitemOpen
  \bibfield  {author} {\bibinfo {author} {\bibfnamefont {A.~B.}\ \bibnamefont
  {Bordo}}, \bibinfo {author} {\bibfnamefont {F.}~\bibnamefont {Gray}}, \ and\
  \bibinfo {author} {\bibfnamefont {D.}~\bibnamefont {Kubiz{\v{n}}{\'{a}}k}},\
  }\href {\doibase 10.1007/JHEP07(2019)119} {\bibfield  {journal} {\bibinfo
  {journal} {JHEP}\ }\textbf {\bibinfo {volume} {07}},\ \bibinfo {pages} {119}
  (\bibinfo {year} {2019})}\BibitemShut {NoStop}%
\bibitem [{\citenamefont {Durka}()}]{Durka:2019}%
  \BibitemOpen
  \bibfield  {author} {\bibinfo {author} {\bibfnamefont {R.}~\bibnamefont
  {Durka}},\ }\href@noop {} {\ }\Eprint {http://arxiv.org/abs/arXiv:1908.04238}
  {arXiv:1908.04238} \BibitemShut {NoStop}%
\end{thebibliography}%

\end{document}